\documentclass[12pt,a4paper]{iopart}

\usepackage{iopams}  
\usepackage{xspace}
\usepackage{graphicx}
\usepackage{dsfont}
\usepackage{color}

\newtheorem{theorem}{Theorem}[section]

\usepackage[bookmarks=true,
            bookmarksopen=true,colorlinks=false]{hyperref}

\newcommand{\R}{\mathds R\xspace}

\newcommand{\Q}{\ensuremath{\mathds Q}\xspace}

\newcommand{\St}{\ensuremath{\mathbb S^2}\xspace}
\newcommand{\Sth}{\ensuremath{\mathbb S^3}\xspace}

\newcommand{\SoXSt}{\ensuremath{\mathbb S^1\!\times \mathbb S^2}\xspace}

\newcommand{\Eqref}[1]{Eq.~\eref{#1}}
\newcommand{\Eqsref}[1]{Eqs.~\eref{#1}}
\newcommand{\Sectionref}[1]{Sec.~\ref{#1}}

\newcommand{\Theoremref}[1]{Theorem~\ref{#1}}

\newcommand{\keyword}[1]{\emph{#1}\xspace}
\newcommand{\eqref}[1]{\eref{#1}}

\newcommand{\ee}{\mathrm e}
\newcommand{\ii}{\mathrm i}
\newcommand{\dd}{\mathrm d}
\newcommand{\E}{\mathcal E}

\newcommand{\p}{_\mathrm{p}}
\newcommand{\f}{_\mathrm{f}}

\newcommand{\eps}{\varepsilon}

\def\Xint#1{\mathchoice
   {\XXint\displaystyle\textstyle{#1}}%
   {\XXint\textstyle\scriptstyle{#1}}%
   {\XXint\scriptstyle\scriptscriptstyle{#1}}%
   {\XXint\scriptscriptstyle\scriptscriptstyle{#1}}%
   \!\int}
\def\XXint#1#2#3{{\setbox0=\hbox{$#1{#2#3}{\int}$}
     \vcenter{\hbox{$#2#3$}}\kern-.5\wd0}}

\def\dashint{\Xint-}

\begin{document}

\title{An exact smooth Gowdy-symmetric generalized Taub-NUT solution}

\author{Florian Beyer and J\"org Hennig}
\address{Department of Mathematics and Statistics,
University of Otago, P.O. Box 56, Dunedin 9054, New Zealand}
\eads{\mailto{fbeyer@maths.otago.ac.nz}, \mailto{jhennig@maths.otago.ac.nz}}

\begin{abstract}
In a recent paper (Beyer and Hennig, 2012 \cite{beyer11}), we have introduced a class of inhomogeneous cosmological models: the smooth Gowdy-symmetric generalized Taub-NUT solutions. Here we derive a three-parametric family of exact solutions within this class, which contains the two-parametric Taub solution as a special case. We also study properties of this solution. In particular, we show that for a special choice of the parameters, the spacetime contains a curvature singularity with directional behaviour that can be interpreted as a ``true spike'' in analogy to previously known Gowdy-symmetric solutions with spatial $\mathbb T^3$-topology. For other parameter choices, the maximal globally hyperbolic region is singularity-free, but may contain ``false spikes''.

\end{abstract}



\section{Introduction}

The fruitful concept of the \keyword{maximal globally hyperbolic development} of Cauchy data was introduced in 1969 by Choquet-Bruhat and Geroch \cite{Choquet1969}. These solutions to Einstein's field equations have, in particular, the property of being uniquely determined (up to isometries) by the Cauchy data on a Cauchy surface.
Sometimes, however, such a maximal globally hyperbolic spacetime can be extended. The extension is not globally hyperbolic and hence there is a \keyword{Cauchy horizon} whose topology and smoothness may in general be very complicated. 
A famous example is the Taub solution \cite{Taub51}, a two-parametric family of spatially homogeneous cosmological models with spatial $\Sth$-topology. This solution can be extended through smooth complete Cauchy horizons with $\Sth$-topology  to the Taub-NUT solutions \cite{NUT63,Misner1963,MisnerTaub1969}. However, there are several non-equivalent extensions \cite{chrusciel93}. Moreover, there exist \emph{closed causal curves} in the extended regions, which is a violation of causality. 

These unexpected properties have raised the question as to whether such pathological phenomena occur only under very special circumstances, like the high symmetry of the Taub solution. Or would quite general solutions in general relativity always suffer from such defects? The former alternative is proposed in Penrose's famous strong cosmic censorship conjecture 
\cite{Penrose69,moncrief81a,Chrusciel91a,Rendall05,Ringstrom09}, according to which the maximal globally hyperbolic development of ``generic'' Cauchy data is inextendible. This means that models like the Taub solution would belong to a (in some sense) negligibly small subset of ``non-generic spacetimes''. However, this hypothesis is far from being proven in the general case.

The interesting features of the Taub-NUT models have motivated the investigation of larger classes of solutions with similar properties. In \cite{moncrief84}, Moncrief studies \keyword{generalized Taub-NUT spacetimes} with a $U(1)$ isometry group and spatial $\Sth$-topology under the assumption of analyticity. Without analyticity, just assuming smoothness, existence of a class of solutions with $U(1)\times U(1)$ symmetry (and again $\Sth$-topology) was shown in \cite{beyer11}. These \keyword{smooth Gowdy-symmetric generalized Taub-NUT solutions} have two functional degrees of freedoms, i.e., for any choice of two smooth functions (subject to a periodicity condition) a corresponding solution exists. {Like the Taub models, they always have a smooth past and (with exception of special singular cases) a smooth future Cauchy horizon} of $\Sth$-topology, through which they can be extended. Properties of these spacetimes have been studied by means of Fuchsian methods and 
soliton methods --- however, without explicitly solving Einstein's vacuum equations. Nevertheless, it is desirable to have exact 
solutions that can be studied in more detail than possible with abstract considerations alone. In this paper we derive and study such an exact solution: a three-parametric, spatially inhomogeneous generalization of the Taub solution. This solution is a particular case of the smooth Gowdy-symmetric generalized Taub-NUT solutions and can be derived with soliton methods. 

The application of methods from soliton theory to the equations of general relativity has a long tradition, in particular for axisymmetric and stationary equilibrium configurations (see, e.g., \cite{BelinskiZakharov1979, KramerNeugebauer1968, KramerNeugebauer1980, Neugebauer1979, NeugebauerMeinel1995, Neugebauer2003, Ruiz1995, Varzugin1997}, to mention just a few of many interesting publications), but also for plane waves and inhomogeneous cosmologies (see, e.g., \cite{AlekseevGriffiths2000,AlekseevGriffiths2004,Lim2008,RendallWeaver2001}). These methods are based on the integrability of the Einstein equations in the case of symmetric spacetimes and make use of reformulations of the nonlinear field equations in terms of associated linear matrix problems. In particular, it is possible to reduce boundary or initial value problems to linear integral equations. Here we will apply a particular approach due to Sibgatullin \cite{Sibgatullin} in order to construct our exact solution.

A particular motivation for a detailed study of the exact solution described in this paper is the following. Since the works of Berger and Moncrief \cite{berger93} on the singularity of Gowdy-symmetric solutions of the vacuum equations with spatial $\mathbb T^3$-topology, there has been increasing evidence that spiky phenomena are a general feature of singular solutions of Einstein's field equations \cite{Lim2009}. While in the $\mathbb T^3$-Gowdy case, solutions with spikes can be ``manufactured'' \cite{RendallWeaver2001,Lim2008} using certain solution generating techniques, the existence and properties of spikes in the case of $\Sth$- or $\SoXSt$-Gowdy solutions is less well understood, in particular due to the degeneracy of the action of the symmetry group at its axes. There are only a few discussions of this in the literature, see \cite{garfinkle99, Stahl02,beyer08}.

{To the best of our knowledge, the family of solutions derived here is the first  example of a family of exact $\Sth$-Gowdy solutions where spiky features develop \textit{on} the symmetry axes.} This is discussed in more detail in \Sectionref{sec:spikes}.

This paper is organized as follows. In Sec.~\ref{sec:Gowdy} we summarize some properties of smooth Gowdy-symmetric generalized Taub-NUT solutions. Then we construct the exact solution in Sec.~\ref{sec:construction}. Afterwards, in Sec.~\ref{sec:prop}, we study various properties of this cosmological model. In particular, we show that the Taub spacetimes are a special case of our solution. Then we look at symmetries of the solution, we show that it is regular in the maximal globally hyperbolic region, and we visualize particular 2-surfaces by embedding them into 3-dimensional Euclidean space. {Moreover, we study singularities that are present for special parameter choices, we extend the solution beyond the Cauchy horizons, and we identify ``false'' and ``true spikes'' on the symmetry axes.} Finally, we discuss our results in Sec.~\ref{sec:discussion}.

\section{Background}
\label{sec:background}

\subsection{Geroch's symmetry reduction and the wave map structure of the vacuum equations}

Without going into the details, let us give a quick summary of Geroch's symmetry reduction \cite{Geroch1} with particular emphasis on the resulting wave map structure of the vacuum field equations; more details which are relevant for our particular case here can be found in \cite{beyer11}. Let $M=\R\times H$ be an oriented and time-oriented
globally hyperbolic $4$-dimensional Lorentzian manifold endowed with a
metric $g_{{a}{b}}$ of signature $(-,+,+,+)$, a smooth global time function $t$ and a Cauchy surface $H$. We denote the chosen volume form associated with $g_{{a}{b}}$ by
$\epsilon_{{a}{b}{c}{d}}$ and the hypersurfaces given by $t=t_0$
for any constant $t_0$ by $H_{t_0}\cong H$. Let $\xi^a$ be a smooth globally defined spacelike Killing
vector field which is tangent to the hypersurfaces $H_t$.
The flow generated by $\xi^a$ induces a map $\pi$ from $M$ to
the space of orbits $S$, i.e., $\pi$ maps every $p\in M$ to the
uniquely determined integral curve of $\xi^a$ starting at
$p$. The quotient manifold $S$ has a canonical smooth structure and is hence a smooth manifold. 

Next, we denote the (square of the) norm of $\xi^a$ by
\begin{equation}
  \label{eq:deflambda}
  \lambda:=g_{ab}\xi^a\xi^b,
\end{equation}
the twist $1$-form of $\xi^a$ as
\begin{equation}
  \label{eq:deftwist}
  \Omega_{a}:=\epsilon_{{a}{b}{c} {d}}\xi^{b}\nabla^{c}\xi^{d},
\end{equation}
where $\nabla$ is the covariant derivative compatible with $g_{ab}$, and the ``$3$-metric'' as
\[h_{{a}{b}}:=g_{{a}{b}}-\frac 1\lambda \xi_{a}\xi_{{b}}.
\] 
 It turns out that there is a unique smooth Lorentzian metric on $S$ which pulls back to
$h_{{a}{b}}$ along $\pi$, which we refer to with the same symbol $h_{ab}$. In the
same way there are a unique function and $1$-form on $S$ which pull back to the function $\lambda$ and the $1$-form $\Omega_{a}$, respectively, on $M$; hence we also denote them by the same symbols.
The quantities $\lambda$,
$\Omega_a$ and $h_{{a}{b}}$ on $S$ completely characterize the local geometry
of $(M,g)$.
Geroch found that Einstein's vacuum field equations on
$(M,g)$ imply that the $1$-form $\Omega_a$ is closed, $\dd\Omega=0$. We therefore locally find a twist potential $\omega$ such that  $\Omega=\dd\omega$.

 Let us define a new smooth Lorentzian metric $\hat h$ on $S$ as
\[\hat h_{{a}{b}}:=\lambda h_{{a}{b}}.\]
We refer to the associated covariant derivative operator as $\hat D_a$,
Ricci tensor as $\hat S_{{a}{b}}$, and raise and lower indices with $\hat h$. Geroch was able to show that the vacuum field
equations for $(M,g)$ (and certain geometric identities) are
equivalent to the following set of equations on $S$:
\begin{eqnarray}
  \label{eq:Gerochevollambda}
  \hat D_a\hat D^a\lambda
  &=\frac 1\lambda\left(\hat D^{a}\lambda\hat D_{a}\lambda
    -\hat D^{a}\omega\hat D_{a}\omega\right),\\
  \label{eq:Gerochevolomega}
  \hat D_a\hat D^a\omega
  &=\frac 2\lambda\hat D^{a}\lambda\hat D_{a}\omega,\\
  \label{eq:GerochRicci3}
  \hat S_{{a}{b}}&=\frac 1{2\lambda^2}\left(
    \hat D_{a}\lambda\hat D_{b}\lambda
    +\hat D_{a}\omega\hat D_{b}\omega\right).
\end{eqnarray}
These equations can be interpreted as $2+1$-dimensional gravity $(S,\hat h)$ coupled to the wave map 
\[u:S\rightarrow\mathcal H,\quad p\mapsto u(p)=(\lambda(p),\omega(p)),\]
 where $\mathcal H$ is the $2$-dimensional half-plane model of the hyperbolic space with coordinates $(\lambda,\omega)$ and metric
\[l=\frac{\dd\lambda^2+\dd\omega^2}{\lambda^2}.\]
The right-hand side of \Eqref{eq:GerochRicci3} can be interpreted as the energy-momentum tensor associated with this wave map.

Given any smooth curve $\gamma(\tau)$ in $S$ and a solution $u=(\lambda,\omega)$ and $\hat h$ of the equations above, the quantity
\begin{equation}
  \label{eq:hypspeed}
s(\tau):=\sqrt{\frac{\left(\frac{\dd}{\dd\tau}(\lambda(\gamma(\tau))\right)^2+\left(\frac{\dd}{\dd\tau}(\omega(\gamma(\tau))\right)^2}{\lambda^2(\gamma(\tau))}}
\end{equation}
is referred to as the \keyword{hyperbolic speed}\footnote{In the literature, it is customary to give this quantity a sign, for example, the same sign as the term $\frac{\dd}{\dd\tau}(\lambda(\gamma(\tau))$ and then to refer to it as  \keyword{hyperbolic velocity}. In this paper, we refrain from doing this.} of $\gamma$.

\subsection{Gowdy-symmetric spacetimes with spatial $3$-sphere topology}
Now, we specialize to the case $H=\Sth$. We
think of \Sth as the submanifold of $\R^4$ determined by
$x_1^2+x_2^2+x_3^2+x_4^2=1$. We are interested in smooth effective
actions of the group $U(1)\times U(1)$ on $\Sth$. From the smooth
effective action of $U(1)$ on $\R^2$ by rotations around the origin, we construct an action of $U(1)\times U(1)$ on $\R^4$ by demanding that the first factor of $U(1)\times U(1)$ generates rotations in the $x_3,x_4=\mathrm{constant}$-planes around the origin, while the second factor generates rotations  in the $x_1,x_2=\mathrm{constant}$-planes around the origin. Clearly, this action is well-defined also when it is restricted to the subset $\Sth$ of $\R^4$. As summarized in \cite{Chrusciel1990}, all smooth effective actions of $U(1)\times U(1)$ are equivalent to this action. It is useful to introduce coordinates $(\theta,\lambda_1,\lambda_2)$ on $\Sth$ so that the $\theta=\mathrm{constant}$-surfaces (wherever they are defined) equal the orbits of the group action. The \keyword{Euler coordinates} $(\theta,\lambda_1,\lambda_2)$ on $\Sth$ are
\begin{eqnarray}
    x_1&=\cos\frac\theta 2\cos\lambda_1,  \quad
    x_2&=\cos\frac\theta 2\sin\lambda_1,\label{eq:euler1}\\
    x_3&=\sin\frac\theta 2\cos\lambda_2,  \quad
    x_4&=\sin\frac\theta 2\sin\lambda_2,\label{eq:euler2}
  \end{eqnarray}
with $\theta\in (0,\pi)$ and $\lambda_1,\lambda_2\in (0,2\pi)$.
Clearly, these coordinates break down at the points $\theta=0$
and $\pi$, which we refer to as ``poles'' or ``axes'' of \Sth in the following.  We also make use of the coordinates
$(\theta,{\rho_1},{\rho_2})$ (which we also call Euler coordinates) with $\theta$ as above and 
\begin{equation}\label{eq:eulerangleparm2}
  \lambda_1=:({\rho_1}+{\rho_2})/2,\quad
  \lambda_2=:({\rho_1}-{\rho_2})/2.
\end{equation}

Let us fix any value $\theta\in[0,\pi]$. Then,
the $2\pi$-periodicity of $\lambda_1$ and $\lambda_2$ implies that, for each choice of $\rho_{1*}\in\R$, all conditions of the form $\rho_1+2\pi k=\rho_{1*}$ given by all integers $k$ yield the same subset of $\Sth$; in the same way,  for each choice of $\rho_{2*}\in\R$, all conditions of the form $\rho_2+2\pi k=\rho_{2*}$ given by all integers $k$ yield the same subset of $\Sth$. In this sense, the coordinates $\rho_1$ and $\rho_2$ are $2\pi$-periodic. However, each of these subsets is a closed curve which is $4\pi$-periodic in $\rho_2$ in the first case and $4\pi$-periodic in $\rho_1$ in the second case.

The coordinate fields $\partial_{\rho_1}$ and
$\partial_{\rho_2}$ (which can be characterized
geometrically without making reference to any coordinates in terms of
left- and right-invariant vector fields of the standard
action of $SU(2)$ on \Sth) are smooth non-vanishing vector fields on \Sth. They are linearly independent everywhere except at the axes where they are parallel: $\partial_{\rho_1}=\partial_{\rho_2}$ at $\theta=0$ and $\partial_{\rho_1}=-\partial_{\rho_2}$ at $\theta=\pi$. The integral curves of both fields are closed circles. The smooth field $\partial_{\lambda_1}$ on the other hand  vanishes at $\theta=\pi$, while $\partial_{\lambda_2}$ vanishes at $\theta=0$.
 Both of the two sets of vector fields span the algebra of generators of the $U(1)\times U(1)$-action on $\Sth$.

Given this action $\Phi: G\times\Sth\rightarrow\Sth$ for $G=U(1)\times U(1)$, we find an action 
\[\tilde\Phi: G\times M\rightarrow M,\quad (u,(t,p))\mapsto (t,\Phi(p)),\] 
where $M=\R\times\Sth$ is equipped with the global smooth time function $t$ above and where $p$ represents any point in $\Sth$. As a consequence the generators of this action are globally defined smooth spacelike vector fields. We assume that we have chosen coordinates $(t,\theta,\rho_1,\rho_2)$ so that $(\theta,\rho_1,\rho_2)$ are Euler coordinates on each $H_t$ and so that $\partial_{\rho_1}$ and $\partial_{\rho_2}$ generate the $U(1)\times U(1)$-action as before.
We shall demand now in addition to the above that $U(1)\times U(1)$ acts \emph{by isometries} on $(M,g)$ and hence that $\partial_{\rho_1}$ and $\partial_{\rho_2}$ span the algebra of Killing vector fields. Let
\begin{equation}
  \label{eq:transformedbasis}
  \xi_1=a \partial_{\rho_1}+b\partial_{\rho_2},\quad \xi_2=c \partial_{\rho_1}+d\partial_{\rho_2}
\end{equation}
be any other two generators of the algebra of Killing vector fields
where $a$, $b$, $c$ and $d$ are real numbers with $ad-bc\not=0$. As long as $a\not=\pm b$ (which is  assumed in all of what follows), it follows that $\xi_1$ never vanishes (in particular not at the axes). This allows us to perform the Geroch reduction with respect to $\xi_1$ globally and hence to define the corresponding projection map $\pi$, quotient manifold $S$, and objects $\lambda$, $\omega$ and $\hat h$ as above. Notice that the twist scalar $\omega$ is defined globally because $M$ and $S$ are simply connected.  
For example, in the case $\xi_1=\partial_{\rho_1}$ (i.e., for $a=1$, $b=0$, and $c,d\in\R$ such that $ad-bc\not=0$), the quotient map is the special Hopf map in \cite{beyer11} and $S=\R\times\St$. Notice that the $1$-parameter subgroup of $U(1)\times U(1)$ generated by $\xi_1$ is not necessarily isomorphic to $U(1)$ since its integral curves are not necessarily closed; indeed, this is the case if and only if $a/b\in\Q$.

Since $[\xi_1,\xi_2]=0$, it follows that the push-forward of $\xi_2$ along $\pi$ is a Killing vector field of $(S,h_{ab})$ and we can perform the Geroch reduction a second time. Since the push-forward of $\xi_2$ vanishes at some points in $S$ (because $\xi_2$ itself must either vanish or must be parallel to $\xi_1$ at some points in $M$), the result is, however, not a smooth manifold, but rather a manifold with boundary. For the discussion in \Sectionref{sec:construction} and in some parts of \Sectionref{sec:prop}, this is not a problem. However, in \Sectionref{sec:spikes}, it is important to use the smooth manifold structure which is obtained by only \emph{one} Geroch reduction with respect to any choice of $\xi_1$ with $a$, $b$, $c$ and $d$ as above.

\subsection{Smooth Gowdy-symmetric generalized Taub-NUT solutions\label{sec:Gowdy}}

In this section, we summarize the definition and some properties of the class of inhomogeneous cosmological models that we have called \keyword{smooth Gowdy-symmetric  generalized Taub-NUT solutions}. For details we refer to \cite{beyer11}.

Such a spacetime is a Gowdy-symmetric, oriented, time-oriented maximally extended globally hyperbolic vacuum solution to Einstein's field equations whose spatial topology is that of the three-sphere $\Sth$. The main property is that it can be extended (not necessarily as a solution of the vacuum equations) to a non-globally hyperbolic Gowdy-symmetric spacetime in the past. The corresponding Cauchy horizon is supposed to be a smooth null surface with $\Sth$-topology and its null generator is parallel to a generator of one of the $U(1)$-factors of the symmetry group on the horizon; in particular, the orbits of the generators are therefore closed.
{In terms of a time function $t\in(0,\pi)$ and the above described coordinates $\theta$, $\rho_1$ and $\rho_2$, one can achieve the following form of the metric \cite{beyer11},} 
\begin{equation}\label{eq:metric}
  g_{ab}=\ee^M(-\dd t^2+\dd\theta^2)+R_0\left[\sin^2\!t\,\ee^u (\dd\rho_1+Q \dd\rho_2)^2+\sin^2\!\theta\,\ee^{-u} \dd\rho_2^2\right],
\end{equation}
where $R_0$ is a positive constant and $u$, $Q$ and $M$ are smooth functions of $t$ and $\theta$. The past Cauchy horizon is located at $t=0$.\footnote{\label{fn:coordinates}{Strictly speaking, the time coordinate $t$ is not defined at $t=0$. However, it is possible to introduce ``regular coordinates'' $(x,y,\rho_1',\rho_2')$ with $x=\cos\theta$ and $y=\cos t$, in which the solution can be extended to and beyond that boundary (see Sec.~\ref{sec:extensions}). The past Cauchy horizon is then located at $y=1$, corresponding to $t=0$. For the sake of simplicity, we will in the following often talk about the surfaces $t=0$ or, in a similar manner, $t=\pi$, without always giving explicit reference to regularized coordinates.}} 
With respect to the choice $\xi_1=\partial_{\rho_1}$ and $\xi_2=\partial_{\rho_2}$ in \Eqref{eq:transformedbasis}, we therefore have
\begin{equation*}
  \lambda=R_0\sin^2\!t\,\ee^u,
\end{equation*}
\begin{equation*}
 \partial_t\omega = -R_0\frac{\sin^3 t}{\sin\theta}\ee^{2u}\partial_\theta Q,\quad
 \partial_\theta\omega = -R_0\frac{\sin^3 t}{\sin\theta}\ee^{2u}\partial_t Q,
\end{equation*}
and
\begin{equation}
  \label{eq:3metric}
  h_{ab}=\ee^M(-\dd t^2+\dd\theta^2)+R_0\sin^2\!\theta\,\ee^{-u} \dd\rho_2^2.
\end{equation}

It was not \emph{a priori} guaranteed that there are any solutions to Einstein's field equations that have all the above properties in the entire time interval {$(0,\pi)$}. In order to establish such a global existence result, we chose the following approach in \cite{beyer11}. In the first step we showed \emph{local} existence in a neighbourhood of the past Cauchy horizon. This was done with the Fuchsian methods developed in \cite{AmesA,AmesB}. With respect to the choice $\xi_1=\partial_{\rho_1}$ and $\xi_2=\partial_{\rho_2}$ in \Eqref{eq:transformedbasis}, the result can be summarized as follows.
\begin{theorem}\label{Thm1}
 Let $S_{**}$ and $Q_{*}$ be axially symmetric functions
  in $C^\infty(\mathbb S^2)$ so that
  $S_{**}(0)=S_{**}(\pi)$
  and $R_0$ a positive constant.  Then there exists a unique smooth
  Gowdy-symmetric generalized Taub-NUT solution for all $t\in(0,\delta]$ (for a sufficiently small $\delta>0$) 
  satisfying the following uniform expansions at $t=0$:
  \begin{eqnarray*}
    R_0\, \ee^{u(t,\theta)} &=& \ee^{S_{**}(\theta)}+O(t^2),\\
    Q(t,\theta)&=&\cos\theta+Q_*(\theta)\sin^2\!\theta 
    +O(t^2),\\
    M(t,\theta)&=&S_{**}(\theta)-2S_{**}(0)+2\ln R_0+O(t^2).
  \end{eqnarray*}
\end{theorem}
Besides local existence, this theorem also shows what the available degrees of freedom are: the two \keyword{asymptotic data functions} $S_{**}$ and $Q_{*}$, which describe the behaviour of the solution in a vicinity of the past horizon.

After local existence on the time interval $(0,\delta]$ was established, we used a global existence result due to Chru\'sciel \cite{Chrusciel1990} that guarantees existence and regularity in the entire time interval $0<t<\pi$. The only remaining question was what happens at $t=\pi$, where the above defined coordinates become singular. In order to answer this question, we applied methods from soliton theory and discussed the linear matrix problem that is equivalent to the essential part of the Einstein vacuum equations under Gowdy symmetry. 
In this way, we were able to find explicit expressions for the metric functions on the boundaries $\theta=0$, $\theta=\pi$ and $t=\pi$ in terms of the data at $t=0$. These expressions were used for an analysis of the solution at $t=\pi$, which strongly indicates the following:
\begin{quote}
 In general, smooth Gowdy-symmetric generalized Taub-NUT solutions (with a
 past Cauchy horizon at $t=0$) develop a second Cauchy horizon
 at $t=\pi$. 
 The only exceptions are special cases in which curvature singularities form. These cases occur when the imaginary part $b=\Im\E$ of the Ernst potential $\E$ [the Ernst potential is defined  in \eqref{eq:deff}-\eqref{eq:defa} below] satisfies
 \begin{equation}
  b_B-b_A= \pm4,
 \end{equation}
 where $b_A=b(t=0,\theta=0)$ and $b_B=b(t=0,\theta=\pi)$ are the values at the poles $A$ and $B$ at $t=0$, see Fig.~\ref{fig:Gowdy} below. Then the solutions have a curvature singularity at $t=\pi, \theta=0$ (for a `$+\!$' sign) or at $t=\pi, \theta=\pi$ (for a `$-\!$' sign), respectively. Hence, whether the solution will be regular or singular at $t=\pi$ can be read off from the data at $t=0$.
\end{quote}
Note, however, that the analysis in \cite{beyer11} does not rule out the possibility that the metric potentials obtained \emph{at} $t=\pi$ do not connect sufficiently smoothly to the potentials at $t<\pi$ (which have not been obtained explicitly). Hence the solutions might develop additional defects as $t\to\pi$, even though we doubt that this can actually happen. Nevertheless, this uncertainty is an additional motivation for studying examples of exact solutions. And the solution presented here turns out to have all expected properties as described above.

Finally, we note that it is assumed in \Theoremref{Thm1} that the past
horizon is generated\footnote{{Note that, strictly speaking, $\partial_{\rho_1}$ is not defined at $t=0$ where the coordinates break down. However, similarly to the remark in footnote~\ref{fn:coordinates}, we can introduce regular coordinates $(x,y,\tilde\rho_1,\tilde\rho_2)$ that extend to the Cauchy horizon. In these coordinates, we find that $\partial_{\tilde\rho_1}=\partial_{\rho_1}$ becomes null on the horizon. Moreover, the integral curves of this vector field are autoparallel curves (i.e.\ ``geodesics in a non-affine parametrization'') which corresponds to a nontrivial surface gravity. Moreover, an appropriate rescaling of the Killing field leads to a vector field whose integral curves are null geodesics. In the following we will nevertheless also refer to $\partial_{\rho_1}$ as a ``generator'' of the Cauchy horizon. The same remark applies to generators of the future Cauchy horizon.}} by the Killing vector field $\partial_{\rho_1}$ and hence its integral curves are closed. In 
the special case $b_A=b_B$, the future horizon is generated by $\partial_{\rho_1}$ as well. Otherwise, the future horizon is generated by $Q\f\partial_{\rho_1}-\partial_{\rho_2}$, where $Q\f$ is the (constant) value of the metric potential $Q$ at the future horizon. In general, this implies that the integral curves of the generator of the future horizon are not closed, except in the special case where $Q\f$ is a rational number. Note also that the metric function $u$ might blow up in the limit $t\to\pi$ even if the spacetime is regular there.

\section{Construction of the exact solution\label{sec:construction}}

\subsection{Einstein's field equations and the Ernst formulation}

The Einstein equations for the metric \eqref{eq:metric} lead to two second-order equations for $u$ and $Q$, which are independent of $M$. Hence one might calculate $u$ and $Q$ in a first step. Afterwards, the remaining Einstein equations provide formulae for $\partial_t M$ and $\partial_\theta M$ so that $M$ can immediately be obtained from a line integral (which turns out to be path independent as a consequence of the field equations for $u$ and $Q$).
The two equations for $u$ and $Q$ are equivalent to the \keyword{Ernst equation}
\begin{equation}\label{eq:EE}
 \Re(\E)\left(-\partial_t^2\E-\cot t\,\partial_t\E
        +\partial_\theta^2\E+\cot\theta\,\partial_\theta\E\right)
 =-(\partial_t\E)^2+(\partial_\theta\E)^2
\end{equation}
for the complex \keyword{Ernst potential} $\E=f+\ii b$, which is constructed from the two Killing vectors $\partial_{\rho_1}$ and $\partial_{\rho_2}$. The real part $f$ of $\E$ is defined by
\begin{equation}\label{eq:deff}
 f:=\frac{1}{R_0}g(\partial_{\rho_2},\partial_{\rho_2})
 =Q^2\ee^u\sin^2\! t+\ee^{-u}\sin^2\!\theta
\end{equation}
and the imaginary part $b$ is given by
\begin{equation}\label{eq:defb}
 \partial_t a=\frac{1}{f^2}\sin t\sin\theta\,\partial_\theta b,\quad
 \partial_\theta a=\frac{1}{f^2}\sin t\sin\theta\,\partial_t b
\end{equation}
with
\begin{equation}\label{eq:defa}
 a:= \frac{g(\partial_{\rho_1},\partial_{\rho_2})}
          {g(\partial_{\rho_2},\partial_{\rho_2})}
   = \frac{Q}{f}\ee^u\sin^2 t.
\end{equation}
Note that the Ernst equation was originally formulated in the context of \emph{axisymmetric and stationary} spacetimes \cite{Ernst1968,KramerNeugebauer1968}. These are characterized by the existence of a space\-like Killing vector (corresponding to axisymmetry) and a second Killing vector (corresponding to stationary), which is timelike in a vicinity of spatial infinity. Since the Gowdy-symmetric solutions also admit two Killing vectors (which, however, both are spacelike), the mathematical formulation of the field equations and the solution methods are very similar in these two cases. Indeed, we may even use the formal coordinate transformation
\begin{equation}\label{eq:transform}
 \rho  = \ii \sin t\sin\theta,\quad
 \zeta = \cos t\cos\theta
\end{equation}
to coordinates ($\rho,\zeta,\rho_1,\rho_2$) in which the metric \eref{eq:metric} takes the Weyl-Lewis-Papapetrou form for axisymmetric and stationary spacetimes. (The two Killing variables $\rho_1$, $\rho_2$ would then play the role of an azimuthal angle and a stationary time coordinate.)

\begin{figure}\centering
 \includegraphics[scale=0.7]{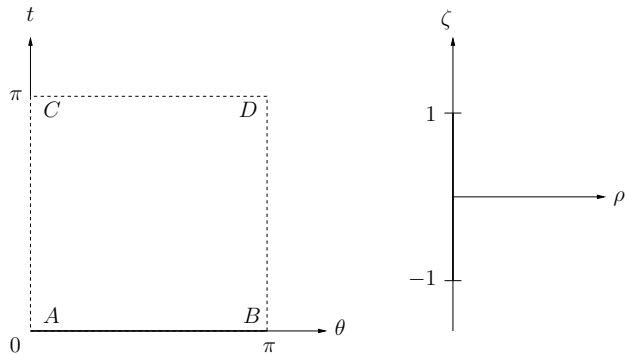}
 \caption{Illustration of an initial value problem for the Ernst equation of a Gowdy-symmetric generalized Taub-NUT solution (with initial data at $t=0$, left panel) and a boundary value problem for the axisymmetric and stationary Ernst equation (with boundary values in the interval $[-1,1]$ on the $\zeta$-axis, right panel). \label{fig:Gowdy}}
\end{figure}

In the following, we wish to solve an \emph{initial value problem} for the Ernst equation \eref{eq:EE}, where we prescribe the initial Ernst potential at $t=0$. However, in terms of the corresponding axisymmetric and stationary formulation, we obtain a \emph{boundary value problem} with prescribed axis values at $\rho=0$, $\zeta\in[-1,1]$ [cf.~\eref{eq:transform}] as illustrated in Fig.~\ref{fig:Gowdy}. Mathematically, initial and boundary value problems have, of course, completely different properties and we cannot expect to find solutions to arbitrary initial value problems from a discussion of a corresponding boundary value problem. On the other hand, in this paper we consider a particular family of solutions where this procedure can indeed be applied. (In any case, one may check afterwards whether the constructed solution really is a solution to the original time-evolution problem.)

A useful method for tackling axisymmetric and stationary boundary value problems is ``Sibgatullin's integral method'' \cite{Sibgatullin}, which we discuss in the next section. For more details on the axisymmetric and stationary Ernst equation and exact solution methods we refer the reader also to \cite{Neugebauer1996} and \cite{Neugebauer2003}.
\subsection{Solution of the Ernst equation}

As shown by Sibgatullin \cite{Sibgatullin}, a boundary value problem for the Ernst equation of an axisymmetric and stationary spacetime can be reformulated in terms of the linear integral equation
\begin{equation}\label{eq:inteq}
 \dashint_{-1}^1\frac{\mu(\xi;\rho,\zeta)[e(\xi)+\tilde e(\eta)]\,\dd\sigma}{(\sigma-\tau)\sqrt{1-\sigma^2}}=0
\end{equation}
for a complex function $\mu(\xi;\rho,\zeta)$, where $\dashint$ denotes the principal value integral. We can fix a unique solution to this homogeneous problem by imposing the additional constraint
\begin{equation}\label{eq:constraint}
 \int_{-1}^1\frac{\mu(\xi;\rho,\zeta)\,\dd\sigma}{\sqrt{1-\sigma^2}}=\pi.
\end{equation}
Here, we have used the definitions $\xi:=\zeta+\ii\rho\sigma$, $\eta:=\zeta+\ii\rho\tau$ with $\sigma,\tau\in[-1,1]$. The boundary values $\E(\rho=0,\zeta)$ appear in the form of their analytical continuations
\begin{equation}
 e(\xi):=\E(\rho=0,\zeta=\xi),\quad
 \tilde e(\xi):=\overline{e(\bar\xi)},
\end{equation}
where the bar denotes complex conjugation. Once $\mu$ is calculated\footnote{For ease of notation, we will often simply write $\mu$ or $\mu(\xi)$ for $\mu(\xi;\rho,\zeta)$.}, the corresponding Ernst potential can be obtained from
\begin{equation}\label{eq:EP}
 \E(\rho,\zeta)=\frac{1}{\pi}\int_{-1}^1\frac{e(\xi)\mu(\xi)\dd\sigma}{\sqrt{1-\sigma^2}}.
\end{equation}

In the following, we intend to construct a family of generalized Taub-NUT solutions for which the initial Ernst potential is simple enough to allow for an exact solution of the integral equation \eref{eq:inteq}, but which contains enough parameters to describe both the regular solutions (with a second Cauchy horizon at $t=\pi$) and the singular cases (with scalar curvature singularities at the points $C$ or $D$ in Fig.~\ref{fig:Gowdy}), see Sec.~\ref{sec:Gowdy}. 

Before we can choose appropriate initial data, we derive some restrictions on the initial Ernst potential $\E\p=f\p+\ii b\p$ at the past horizon $t=0$ (or, equivalently, on the boundary values at $\rho=0$ in the corresponding boundary value problem).
At $t=0$, the real part $f$ of $\E$ and the regular metric potential $u$ are related by
\begin{equation}
 f(t=0,\theta)=\ee^{-u(t=0,\theta)}\sin\!^2\theta,
\end{equation}
see \eqref{eq:deff}.
As a consequence, $f\p$ has to satisfy the conditions
\begin{equation}\label{eq:condf1}
 f\p(\zeta=\pm1)=0,\quad
 f\p(\zeta)>0\textrm{ for } -1<\zeta<1,
\end{equation}
because $\theta=0,\pi$ corresponds to $\zeta=\pm1$ for $t=0$.
A second restriction on $f\p$ follows from the requirement that the first-order equations for the metric potential $M$ must have a regular solution. This led to the condition $S_{**}(0)=S_{**}(\pi)$ in Theorem~\ref{Thm1}, which translates into
\begin{equation}\label{eq:condf2}
 \frac{\dd f\p}{\dd\zeta}\Big|_{\zeta=1}=-\frac{\dd f\p}{\dd\zeta}\Big|_{\zeta=-1}.
\end{equation}
Finally, a condition for the imaginary part $b\p$ follows from the relation between $b\p$ and the metric potential $Q$ \cite{beyer11},
\begin{equation}
 b\p(\theta)=b_A+2\int_0^\theta Q(0,\theta')\sin\theta'\,\dd\theta',
\end{equation}
where $b_A=b(t=0,\theta=0)$ is the value of $b$ at the point $A$, see Fig.~\ref{fig:Gowdy}. In our setting, the function $Q$ takes on the
boundary values $Q=1$ for $\theta=0$ and $Q=-1$ for $\theta=\pi$. Using $\zeta=\cos(\theta)$ for $t=0$ together with the latter equation, these boundary conditions lead to
\begin{equation}\label{eq:condb}
 \frac{\dd b\p}{\dd\zeta}\Big|_{\zeta=1}=-2,\quad \frac{\dd b\p}{\dd\zeta}\Big|_{\zeta=-1}=2.
\end{equation}

As probably the simplest non-trivial possibility for the initial Ernst potential $\E\p$, we choose a cubic imaginary part $b\p=c_0+c_1\zeta+c_2\zeta^2+c_3\zeta^3$. The constant $c_0$, which plays the role of an integration constant in \eqref{eq:defb}, has no physical meaning. Hence we may set $c_0=0$. Now we must ensure that \eref{eq:condb} holds, which leads to $c_2=-1$ and $c_1=-3c_3$. Thus we arrive at
\begin{equation}\label{eq:indatb}
 b\p(\zeta)=c_3\zeta(\zeta^2-3)-\zeta^2.
\end{equation}
For the real part $f\p$, subject to the conditions \eref{eq:condf1}, \eref{eq:condf2}, we choose a quadratic function 
\begin{equation}\label{eq:indatf}
 f\p=c_1(1-\zeta^2) 
\end{equation}
with $c_1>0$ (which is not related to the auxiliary quantity $c_1$ above).
However, for our choice of a cubic function $b\p$, it turns out that the method for solving the integral equation \eref{eq:inteq} as  described in the following  will only work if $f\p$ is a cubic polynomial, too. A possible way out is to start from the \emph{cubic} initial potential
\begin{equation}\label{eq:indat}
 \E\p=c_1(1-\zeta^2)\left(1-\frac{\zeta}{d}\right)
      +\ii\zeta\left[c_3(\zeta^2-3)-\zeta\right]
\end{equation}
depending on the two parameters $c_1$ and $c_3$ and on an auxiliary parameter $d$. At the end, when we have constructed $\E$, we may take the limit $d\to\infty$ in which the real part of \eref{eq:indat} reduces to \eref{eq:indatf}. (Note that the condition \eqref{eq:condf2} is only satisfied in the limit $d\to\infty$, i.e.\ a finite $d$ cannot lead to a regular solution of our original initial value problem.)

Now we have to solve the integral equation \eqref{eq:inteq} for our choice \eqref{eq:indat} of the initial potential. According to \cite{Sibgatullin}, it is not too difficult to find exact solutions of \eqref{eq:indat} for \emph{rational} initial data. In that case, one needs to find the zeros $\xi_1,\dots,\xi_N$ of the equation $e(\xi)+\tilde e(\xi)=0$ together with their multiplicities $m_1,\dots,m_N$. The solution $\mu(\xi)$ should then have the form
\begin{equation}\label{eq:ansatz}
 \mu(\xi)=A(\rho,\zeta)+\sum\limits_{k=1}^N
  \left[\frac{A^1_k(\rho,\zeta)}{\xi-\xi_k}
       +\frac{A^2_k(\rho,\zeta)}{(\xi-\xi_k)^2}+\dots 
       +\frac{A^{m_k}_k(\rho,\zeta)}{(\xi-\xi_k)^{m_k}}\right].
\end{equation}
The unknown functions $A$ and $A^n_k$ can be found from the algebraic system of equations that one obtains by plugging \eqref{eq:ansatz} into \eqref{eq:inteq}, \eqref{eq:constraint}.

In our case we have to solve the equation
\begin{equation}
 e(\xi)+\tilde e(\xi)\equiv 2c_1(1-\xi^2)\left(1-\frac{\xi}{d}\right)=0,
\end{equation}
which has the solutions $\xi=\pm 1,d$ (of respective multiplicities one). Hence we start from the ansatz
\begin{equation}
 \mu(\xi)=A(\rho,\zeta)+\frac{A_+(\rho,\zeta)}{\xi+1}+\frac{A_-(\rho,\zeta)}{\xi-1}+\frac{A_d(\rho,\zeta)}{\xi-d}.
\end{equation}
In order to determine the functions $A$, $A_{\pm}$ and $A_d$ we need to evaluate the integrals in \eqref{eq:inteq} and \eqref{eq:constraint}, which can be done with the aid of the formulae
\begin{equation}\fl
 \int_{-1}^1\frac{\dd\sigma}{\sqrt{1-\sigma^2}}=\pi,\quad
 \int_{-1}^1\frac{\xi\dd\sigma}{\sqrt{1-\sigma^2}}=\pi\zeta,\quad
 \int_{-1}^1\frac{\xi^2\dd\sigma}{\sqrt{1-\sigma^2}}=\pi\left(\zeta^2-\frac{\rho^2}{2}\right)
\end{equation}
and
\begin{equation}\fl
 \dashint_{-1}^1\frac{\dd\sigma}{\sqrt{1-\sigma^2}(\sigma-\tau)}=0,\quad
 \int_{-1}^1\frac{\dd\sigma}{\sqrt{1-\sigma^2}(\xi-\alpha)}=\frac{\pi\,\mathrm{sgn}(\zeta-\alpha)}{\sqrt{\rho^2+(\zeta-\alpha)^2}}\quad\textrm{for}\quad \alpha\in\R.
\end{equation}
As the first step, we find that the constraint \eqref{eq:constraint} leads to
\begin{equation}\label{eq:con0}
 A+\frac{A_+}{r_+}-\frac{A_-}{r_-}-\frac{A_d}{r_d}=1
\end{equation}
for $\zeta\in[-1,1]$, $d>1$, where 
\begin{equation}
 r_\pm:=\sqrt{\rho^2+(\zeta\pm1)^2},\quad
 r_d:=\sqrt{\rho^2+(\zeta-d)^2}.
\end{equation}
 Similarly, we obtain from \eqref{eq:inteq} that
\begin{equation}\label{eq:con1}
 T_0+\zeta T_1+\left(\zeta^2-\frac{\rho^2}{2}\right)T_2+\frac{T_+}{r_+}-\frac{T_-}{r_-}-\frac{T_d}{r_d}=0,
\end{equation}
where
\begin{eqnarray*}\fl
 T_0 & = & -\left(\frac{c_1}{d} + 3 \ii c_3\right) A 
           - \left[\left(\frac1 d + 1\right) c_1 + \ii (c_3 + 1)\right] A_+\\ \fl 
     &&    + \left[\left(\frac1 d - 1\right) c_1 
           +    \ii (c_3 - 1)\right] A_-
         + \ii (c_3 d - 1) A_d \\ \fl
     &&    - \left[(\ii + c_1) A - \left(\frac{c_1}{d} + \ii c_3\right) 
             (A_+ + A_- + A_d)\right] \eta
        + \left(\frac{c_1}{d} + \ii c_3\right) A \eta^2,\\ \fl
 T_1 & = &  -(\ii + c_1) A + \left(\frac{c_1}{d} + \ii c_3\right) (A_+ + A_- + A_d)
          + \left(\frac{c_1}{d} + \ii c_3\right) A \eta,\\ \fl
 T_2 & = & \left(\frac{c_1}{d} + \ii c_3\right) A,\\ \fl
 T_+ & = & -\left[[c_1 + \ii (2 c_3 - 1)] - \left(\left(\frac1 d + 1\right) c_1 - \ii (c_3 + 1)\right) \eta + \left(\frac{c_1}{d} - \ii c_3\right) \eta^2\right] A_+,\\ \fl
 T_- & = & \left[[c_1 - \ii (2 c_3 + 1)] + \left(\left(-\frac1 d + 1\right) c_1 + \ii (c_3 - 1)\right) \eta - \left(\frac{c_1}{d} -\ii c_3\right) \eta^2\right] A_-,\\ \fl
 T_d & = & \left[\left(\frac{c_1}{d} - \ii (3 c_3 + d - c_3 d^2)\right) - 
   \ii (1 - c_3 d) \eta - \left(\frac{c_1}{d} - \ii c_3\right) \eta^2\right] A_d.
\end{eqnarray*}
The left hand side of \eqref{eq:con1}, which is  quadratic in $\eta$, must vanish for all $\eta$. Hence, by separately equating the coefficients of $\eta^0$, $\eta^1$ and $\eta^2$ to zero, we find three further algebraic equations. Together with \eqref{eq:con0}, we arrive at a system of four algebraic equations for the four unknowns $A, A_\pm, A_d$. It is a lengthy but straightforward calculation to solve this system and to plug the solution into formula \eqref{eq:EP} for the Ernst potential. Afterwards, we can proceed with our programme and take the limit $d\to\infty$. (As explained above, the parameter $d$ is only an auxiliary quantity introduced for technical reasons. At the end, however, we are only interested in the Ernst potential in the limit $d\to\infty$.)

In a next step we ``transform'' the obtained solution of the axisymmetric and stationary Ernst equation into a solution of our original time-evolution problem by virtue of the coordinate transformation \eqref{eq:transform}. In particular, we replace $r_+$ with $\cos t+\cos\theta$ and $r_-$ with $\cos t-\cos\theta$. In this way, we arrive at the following Ernst potential,
\begin{eqnarray}\fl
 \E & = & -\Big\{c_3^4 (x^2-1)^3 (y-1)^6 (y+1) + 
    16 c_1^4 (x^2-1) (y+1)^3   -  16 \ii c_1^3 (y+1) \Big[y^2 + 4 y-5
   \nonumber\\ \fl
    && \quad  - x^2 (y^2 - 4 y+7) - 
       c_3 x \Big(y^3+ 2 y^2-y  + 10   - x^2 (y^3+2y^2-y+2)\Big)\Big]
   \nonumber\\ \fl
    &&   
    \quad +4 \ii c_1 (x^2-1) (y-1)^3 \Big[4 - 4 c_3 x (3 y+2) + 
       c_3^2 \Big(y^2+4 y  +19 + x^2 (11 y^2+20 y +9)\Big)
   \nonumber\\ \fl
    &&  \quad + 
       c_3^3 x \Big(y^3 + 10 y^2+ 31 y +18  + 
          x^2 (3 y^3+ 14 y^2+ 17 y+6)\Big)\Big]
   \nonumber\\ \fl
    && \quad + 
      4 c_1^2 (y-1) \Big[4 \Big(- y^2+ 4 y-3  + x^2 ( y^2+4 y+7)\Big) 
   \nonumber\\ \fl
    && \quad + 8 c_3 x \Big( y^3 + 2 y^2+ 5 y +10  - 
          x^2 (y^3 + 6 y^2+ 7 y  +2)\Big) 
   \nonumber\\ \fl
    && \quad + 
       c_3^2 \Big( 3 y^4+ 8 y^3 + 26 y^2+ 56 y  +51 - 
          2 x^2 (5 y^4 + 16 y^3 + 32 y^2 + 24 y-5 )
   \nonumber\\ \fl 
    && \quad + 
          x^4 (7 y^4 + 24 y^3 + 38 y^2+ 24 y +3
             )\Big)\Big]\Big\}
    \nonumber\\ \fl
    &&\quad /\Big\{16 c_1 \Big[c_3^2 (x^2-1) (y-1)^3 + 
      4 c_1^2 (y+1)
     + 4 \ii c_1 (y-1) \Big(1 - c_3 x (y+2)\Big)\Big]\Big\}\label{eq:E},
\end{eqnarray}
where $x:=\cos\theta$, $y:=\cos t$. One can explicitly verify that the Ernst equation \eqref{eq:EE} is satisfied for this potential, i.e.\ we have indeed constructed the Ernst potential of a smooth Gowdy-symmetric generalized Taub-NUT solution.

\subsection{Metric potentials}
\label{sec:metricpotentials}
In order to obtain the corresponding metric potentials $u$, $Q$ and $M$, we could proceed as follows. In a first step, we calculate the auxiliary quantity $a$ from $\E$ via line integration using \eqref{eq:defb}. Then we solve \eqref{eq:deff}, \eqref{eq:defa} for $u$ and $Q$ to obtain
\begin{equation}\label{eq:uQ}
 \ee^u=\frac{f a^2}{\sin^2\! t}+\frac{\sin^2\!\theta}{f},\quad
 Q=\frac{f^2 a}{f^2 a^2 + \sin^2\! t\sin^2\!\theta}.
\end{equation}
Finally, we compute $M$ from a line integral using \cite{beyer11}
\pagebreak
\begin{eqnarray}\fl\label{eq:M1}
 (\cos^2\! t-\cos^2\!\theta)\partial_t M & = &
  \frac{1}{2}\ee^{2u}\frac{\sin^3\! t}{\sin\theta}
 \Big[\cos t\sin\theta[(\partial_t Q)^2+(\partial_\theta Q)^2]
       -2\sin t\cos\theta (\partial_t Q)(\partial_\theta Q)\Big]
 \nonumber\\
 & & +\frac{1}{2}\sin t \sin\theta
 \Big[\cos t\sin\theta[(\partial_t u)^2+(\partial_\theta u)^2]
       -2\sin t\cos\theta (\partial_t u)(\partial_\theta u)\Big]
 \nonumber\\
 & & -(2\cos^2\!t\,\cos^2\!\theta\,-\cos^2\!t-\cos^2\!\theta)
      \,\partial_t u
 \nonumber\\
 & & -2\sin t\cos t\sin\theta\cos\theta(\partial_\theta u+\tan\theta),  
\end{eqnarray}
\begin{eqnarray}\label{eq:M2}\fl
 (\cos^2\! t-\cos^2\!\theta)\partial_\theta M & = &
  -\frac{1}{2}\ee^{2u}\frac{\sin^3\! t}{\sin\theta}
 \Big[\sin t\cos\theta[(\partial_t Q)^2+(\partial_\theta Q)^2]
       -2\cos t\sin\theta (\partial_t Q)(\partial_\theta Q)\Big]
 \nonumber\\
 & & -\frac{1}{2}\sin t \sin\theta
 \Big[\sin t\cos\theta[(\partial_t u)^2+(\partial_\theta u)^2]
       -2\cos t\sin\theta (\partial_t u)(\partial_\theta u)\Big]
 \nonumber\\ 
 & & -2\sin t\cos t\sin\theta\cos\theta(\partial_t u-\tan t)
 \nonumber\\
 & & -(2\cos^2\!t\,\cos^2\!\theta\,-\cos^2\!t-\cos^2\!\theta)
      \,\partial_\theta u.
\end{eqnarray}
However, it turns out that the first step, i.e.\ the calculation of $a$ from \eqref{eq:defb}, leads to fairly complicated integrals which cannot easily be solved. Fortunately, as an alternative to \eqref{eq:defb}, the function $a$ may also be calculated directly from the solution $\mu(\xi)$ of the integral equation \eqref{eq:inteq}. As shown by Manko and Sibgatullin \cite{Manko93}\footnote{The quantity $\omega$ in Eq. (3.21) of \cite{Manko93} is the negative of our function $a$.}, $a$ is given by
\begin{equation}
 a=\frac{2}{\pi f}\,\Im\int_{-1}^1\frac{\xi\mu(\xi)\,\dd\sigma}{\sqrt{1-\sigma^2}}.
\end{equation}
Applying this formula to $\mu(\xi)$ as given in \eqref{eq:ansatz}, we obtain
\begin{equation}
 a=\frac{2}{f}\Im\left[\zeta A+\left(1-\frac{1}{r_+}\right)A_++\left(1-\frac{1}{r_-}\right)A_-+\left(1-\frac{1}{r_d}\right)A_d\right].
\end{equation}
Here, we can replace $A$, $A_\pm$, $A_d$ by the solutions of the algebraic system of equations as discussed above. Afterwards, we again take the limit $d\to\infty$ and then transform the solution to the coordinates $t$, $\theta$ via \eqref{eq:transform}. In this way, we obtain the correct function $a$ for our time-evolution problem. 

Finally, we may calculate $u$ and $Q$ from $a$ and $f$ using \eqref{eq:uQ}. The results are the remarkably simple functions
\begin{eqnarray}\fl
 \ee^u & = & 16 c_1 [c_3^2 (1-x^2) (1-y)^3 + 4 c_1^2 (1 + y)]
           /\Big[(1 + 
   y) \Big(c_3^4 (1-x^2)^2 (1-y)^6 + 16 c_1^4 (1+y)^2
   \nonumber\\ \fl
   && + 
   8 c_1^2 (1-y)^2 [2 - 4 c_3 x (y+2)
       + 
      c_3^2 (1 - y^2 + x^2 (3 y^2+ 8 y+7))]\Big)\Big],\label{eq:solu}\\ \fl
   Q & = & x+\frac{c_3}{8} (1-x^2) \Big[4 c_1^2 (y^3 + 5 y^2+ 11 y +7 )+(1-y)^3 \Big(4 - 
      8 c_3 x (y+2) 
    \nonumber\\ \fl
     &&  + 
      c_3^2 [y^2+ 4 y +7 + 3 x^2 (y^2 + 4 y+3 )]\Big)\Big]
     /[c_3^2 (1-x^2) (1-y)^3 + 4 c_1^2 (1+y)],\label{eq:solQ}
\end{eqnarray}
where we again used the abbreviations $x=\cos\theta$, $y=\cos t$.
From these expressions for $u$ and $Q$, we may calculate the remaining metric potential $M$ with \eqref{eq:M1}, \eqref{eq:M2}. The corresponding integration can   be done explicitly and we obtain
\begin{eqnarray}\fl
 \ee^M & = & c \Big[c_3^4 (x^2-1)^2 (y-1)^6 + 16 c_1^4 (y+1)^2 + 
  8 c_1^2 (y-1)^2 \Big(2 - 4 c_3 x (y+2)
  \nonumber\\ \fl
  && + 
     c_3^2 [1 - y^2 + x^2 (3 y^2+8 y +7)]\Big)\Big],\label{eq:solM}
\end{eqnarray}
where $c>0$ is an integration constant. However, $c$ cannot be chosen freely but is fixed by axis regularity conditions. It follows from the analysis in \cite{beyer11} that a combination of the potentials $M$ and $u$ must be constant on the axes,
\begin{equation}
 \theta=0,\pi:\quad \ee^{M+u}=R_0.
\end{equation}
For the above functions $u$ and $M$ we find $\lim_{\theta\to0/\pi}\ee^{M+u}=64cc_1^3$. Thus $c$ is given by
\begin{equation}\label{eq:c}
 c=\frac{R_0}{64c_1^3}.
\end{equation}

We have now found all metric potentials corresponding to our initial data \eqref{eq:indatb}, \eqref{eq:indatf} and in this way
constructed a family of smooth Gowdy-symmetric generalized Taub-NUT solutions depending on the three parameters $c_1>0$, $c_3\in\R$ and $R_0>0$.

Finally, we note that the metric potentials can be written in the concise form
\begin{eqnarray}\label{eq:solnew1}
 \ee^M & = & \frac{R_0}{64 c_1^3}(U^2+V^2),\quad
 \ee^u = \frac{R_0 }{4c_1^2}\cdot\frac{U\ee^{-M}}{1+y},\\
 \label{eq:solnew2}
 Q & = & x+\frac{c_3}{8}(1-x^2)\left(7+4y+y^2+\frac{(1-y)V^2}{4c_1^2U}\right)
\end{eqnarray}
with
\begin{equation}\fl\label{eq:defUV}
 U  := c_3^2(1-x^2)(1-y)^3+4c_1^2(1+y),\quad
 V  :=  4c_1(1-y)[1-c_3x(2+y)].
\end{equation}

\section{Properties of the solution\label{sec:prop}}

\subsection{Taub solution\label{sec:Taub}}

The solution derived above contains the Taub solution \cite{Taub51} as a special case. If we set
\begin{equation}
 c_3=0 
\end{equation}
and replace the parameters $c_1$ and $R_0$ in terms of constants $l$ and $m$ via
\begin{equation}
 c_1=\frac{1}{l}\left(\sqrt{l^2+m^2}+m\right),\quad
 R_0=2l\sqrt{l^2+m^2},
\end{equation}
then the solution \eqref{eq:solnew1}-\eqref{eq:solnew2} 
simplifies to
\begin{equation}
  \label{eq:taubsolutions}
 \ee^M=l^2+\left(m+\sqrt{l^2+m^2}y\right)^2,\quad
 \ee^u=2l\sqrt{l^2+m^2}\,\ee^{-M},\quad
 Q=x.
\end{equation}
This is indeed the Taub solution in our coordinates \eqref{eq:metric}, see \cite{beyer11}.

\subsection{Discrete symmetry}

It follows immediately from \eqref{eq:solu}-\eqref{eq:solM} that $u$ and $M$ are invariant under the transformation
\begin{equation}
 c_3\mapsto -c_3, \quad
 \theta\mapsto\pi-\theta \quad (\Leftrightarrow x\mapsto -x),
\end{equation}
whereas $Q$ changes into $-Q$. As a consequence, we see that the metric \eqref{eq:metric} is invariant under the simultaneous transformation
\begin{equation}\label{eq:sym}
 c_3\mapsto -c_3,\quad
 \theta\mapsto \pi-\theta,\quad
 \rho_2\mapsto -\rho_2,
\end{equation}
which interchanges the axes $\theta=0$ and $\theta=\pi$.

\subsection{Regularity\label{sec:regularity}}

As discussed in Sec.~\ref{sec:Gowdy}, smooth Gowdy-symmetric generalized Taub-NUT solutions are regular for $t\in (0,\pi)$ and they can be smoothly extended through $t=0$. Moreover, it is expected that they can also be smoothly extended through $t=\pi$, provided $b_B-b_A\neq \pm4$ holds, where $b_A=b(t=0,\theta=0)$ and $b_B=b(t=0,\theta=\pi)$. If this condition is violated, then we expect scalar curvature singularities at the points $C$ or $D$, see Fig.~\ref{fig:Gowdy}. For our solution \eqref{eq:solu}-\eqref{eq:solM}, we find $b_A=-1-2c_3$, $b_B=-1+2c_3$ and hence
\begin{equation}
 b_B-b_A=4c_3.
\end{equation}
Consequently, the solution should be regular as long as $c_3\neq\pm1$.
That this is true can easily be verified by calculating the components $g_{ij}$ of the metric in terms of the functions $\ee^M$, $\ee^u$ and $Q$ in \eqref{eq:solnew1}, \eqref{eq:solnew2}. All components turn out to be analytic functions of $x=\cos\theta$ and $y=\cos t$ everywhere in the interior of the Gowdy square, i.e.\ for $\theta\in(0,\pi)$, $t\in(0,\pi)$, provided that $\ee^M\neq0$ holds. 
Moreover, the determinant of the metric is
\begin{equation}\label{eq:det}
 \det(g)=-R_0^2\,\ee^{2M}\sin^2\!\theta\,\sin^2\! t,
\end{equation}
i.e.\ the metric is non-degenerate in the interior of the Gowdy square, again under the condition $\ee^M\neq 0$. (Note, however,  
that the above representation of the metric in terms of Euler coordinates \emph{is} degenerate at the boundaries $\theta=0,\pi$ and $t=0,\pi$ as a consequence of coordinate singularities. 
At the axes $\theta=0,\pi$ we find the usual axes singularities, which can be removed by locally introducing Cartesian coordinates\footnote{In \ref{App2}, where geodesics at $\theta=0$ and $\theta=\pi$ are calculated, it is shown how the axes singularities can be removed.}. And the coordinate singularities at the Cauchy horizons at $t=0$ and $t=\pi$  can also be removed by introducing suitable ``regular'' coordinates, see Sec.~\ref{sec:extensions} below.)

We conclude from the above discussion that the regularity of the solution is related to the zeros of $\ee^M$.
In order to find out whether $\ee^M$ can vanish for $x\in[-1,1]$, $y\in[-1,1]$, we note that according to \eqref{eq:solnew1}, $\ee^M=0$ is equivalent to $U=V=0$. This leads to the conditions
\begin{equation}
 y+1=0,\quad
 1-x^2=0,\quad
 1-c_3(y+2)x=0 
\end{equation}
for vanishing $\ee^M$, which have the two solutions
\begin{eqnarray}
 c_3=\pm1,\quad x=c_3,\quad y=-1.
\end{eqnarray}
This shows that $\ee^M$ can never vanish in the \emph{interior} of the Gowdy square, but in the singular cases $c_3=\pm1$, there are zeros at the \emph{boundary points} $C$ ($x=1$, $y=-1$) or $D$ ($x=-1$, $y=-1$), respectively.

We may also calculate the Kretschmann scalar $K=R_{ijkl}R^{ijkl}$, which turns out to have the form
\begin{equation}\label{eq:Kret}
 K(x,y)=\frac{P(x,y)}{\ee^{6M(x,y)}},
\end{equation}
where $P$ is a lengthy polynomial in $x$ and $y$. Obviously, also $K$ is regular wherever $\ee^M\neq0$ holds. 
Hence we conclude that the Kretschmann scalar is bounded in the entire Gowdy square --- with exception of the two singular cases $c_3=\pm1$, in which $K$ diverges as expected at the points $C$ or $D$.

\subsection{Embedding of 2-surfaces}

In order to get a better idea of the geometric properties of the solution, it is interesting to visualize particular 2-surfaces by embedding them in Euclidean space. To find appropriate 2-surfaces\footnote{Suppose that we have at least one spacelike Killing vector field $\xi$. Then a more geometrical construction of 2-spheres on the basis of our discussion in \Sectionref{sec:background} (more details are given in \cite{beyer11}) is as follows. Since the Hopf map maps $M$ to the quotient manifold $\R\times\St$ with a natural $2+1$-dimensional Lorentzian metric,  the $t=\mathrm{constant}$-surfaces in the quotient manifold are naturally homeomorphic to $\St$ and their induced metric is Riemannian. A comparison with \Eqref{eq:3metric} reveals that this metric can be expressed explicitly in terms of the function $u$ and $M$ in analogy to \Eqref{eq:metrich}. In the Gowdy case, the result depends on the choice of the Killing vector field $\xi=\xi_1$.}, we start by considering the embedding of the 3-sphere \Sth in 
$\
R^4$ with Euler coordinates \eqref{eq:euler1}, \eqref{eq:euler2},
\begin{eqnarray}\label{eq:emb1}
 x_1=\cos\frac\theta 2 \cos\lambda_1,\quad 
 x_2=\cos\frac\theta 2 \sin\lambda_1,\\
 \label{eq:emb2}
 x_3=\sin\frac\theta 2 \cos\lambda_2,\quad
 x_4=\sin\frac\theta 2 \sin\lambda_2.
\end{eqnarray}
Here, $x_1,\dots,x_4$ are Cartesian coordinates in $\R^4$ and $\theta$, $\lambda_1$, $\lambda_2$ are coordinates in \Sth. The relation between $\lambda_1$, $\lambda_2$ and our coordinates $\rho_1$, $\rho_2$ is [cf.~\eqref{eq:eulerangleparm2}]
\begin{equation}
 \lambda_1=\frac{\rho_1+\rho_2}{2},\quad
 \lambda_2=\frac{\rho_1-\rho_2}{2}.
\end{equation}
It follows from \eqref{eq:emb1}, \eqref{eq:emb2} that $\lambda_2=0$ is a two-dimensional hemisphere (with $x_3\ge0$ and such that $\theta=\pi$ corresponds to the north pole and $\theta=0$ to the equator) in the three-dimensional space $x_4=0$. Similarly, the subspace $\lambda_2=\pi$ describes a hemisphere with $x_3\le 0$ (where the south pole and the equator are obtained for $\theta=\pi$ and $\theta=0$, respectively). Hence, a complete 2-sphere can be obtained by considering $\lambda_2=0$ and $\lambda_2=\pi$ together, which corresponds to $\rho_1=\rho_2$ and $\rho_1=\rho_2+2\pi$.

Since slices $t=\textrm{constant}$ of smooth Gowdy-symmetric generalized Taub-NUT solution have \Sth-topology, we may expect with the above discussion that subspaces
\begin{equation}
 \Sigma=\Sigma_1\cup\Sigma_2,\quad
\end{equation}
with
\begin{eqnarray}
 \Sigma_1:=\{\theta\in[0,\pi], t=t_0, \rho_1=\rho_2\in[0,2\pi)\},\\
 \Sigma_2:=\{\theta\in[0,\pi], t=t_0, \rho_1=\rho_2+2\pi\in[0,2\pi)\},
\end{eqnarray}
describe two-dimensional surfaces with \St-topology for any $t_0\in(0,\pi)$.

In the following we try to embed $\Sigma$ isometrically into $\R^3$. It is generally not guaranteed that such an embedding exists globally, but we will see that this is possible for some surfaces $\Sigma$.
To this end we set $\dd t=0$ and $\dd\rho_1=\dd\rho_2=:\dd\varphi$ in \eqref{eq:metric} to obtain the metric $h$ in $\Sigma$,
\begin{equation}\label{eq:metrich}
 h=\ee^M\dd\theta^2+R_0[\sin^2\!\theta\,\ee^u(1+Q)^2+\sin^2\!\theta\,\ee^{-u}]\dd\varphi^2.
\end{equation}
In a next step, we perform a coordinate transformation $\theta=\theta(\alpha)$ and investigate whether the metric $h$ in these coordinates can be brought to the form
\begin{equation}
 h=(r^2+r'^{\,2})\dd\alpha^2+r^2\sin^2\!\alpha\,\dd\varphi^2
\end{equation}
for an appropriate function $r=r(\alpha)$ describing the embedded surface in spherical coordinates $(r,\alpha,\varphi)$, where a prime $'$ denotes a derivative with respect to $\alpha$. Hence we have to solve the two equations
\begin{equation}\label{eq:emb}
 r^2+r'^{\,2}=\ee^M\theta'^{\,2},\quad
 r^2\sin^2\!\alpha=R_0[\sin^2\!\theta\,\ee^u(1+Q)^2+\sin^2\!\theta\,\ee^{-u}],
\end{equation}
which can be done numerically\footnote{For a numerical solution, we use the second equation in \eqref{eq:emb} to eliminate $r$ and $r'$ from the first equation. This leads to an ODE of the form $\theta'(\alpha)=F(\theta,\alpha)$. Starting from the north pole $\alpha=0$, where we have the initial condition $\theta=\pi$ according to the above discussion, we solve the ODE with a fourth-order Runge-Kutta method until $\theta=0$ is reached (corresponding to the equator). This provides the upper ``hemisphere'' of the embedded figure --- the lower one is obtained from a reflection. A technical detail is a degeneracy of the equation at $\alpha=0$, which allows to choose the initial derivative $\theta'$  in addition to the function value, where the particular value $\theta'(0)$ is unimportant and just fixes the origin of the polar coordinates. At the end we shift the embedding diagram to obtain a symmetric picture.}. It turns out that the embedding in $\R^3$ for surfaces $\Sigma$ near $t=\pi$ is only possible for 
negative values of $c_3$. On the other hand, we could consider slices $\lambda_1=0,\pi$ instead of $\lambda_2=0,\pi$, in which case embeddings for positive $c_3$ were possible. However, because of the invariance of the solution under the transformation \eqref{eq:sym}, it is sufficient to consider $c_3\le 0$, which we will do in the remainder of this subsection.

\begin{figure}\centering
 \vspace{2mm}
 \includegraphics[scale=1.0]{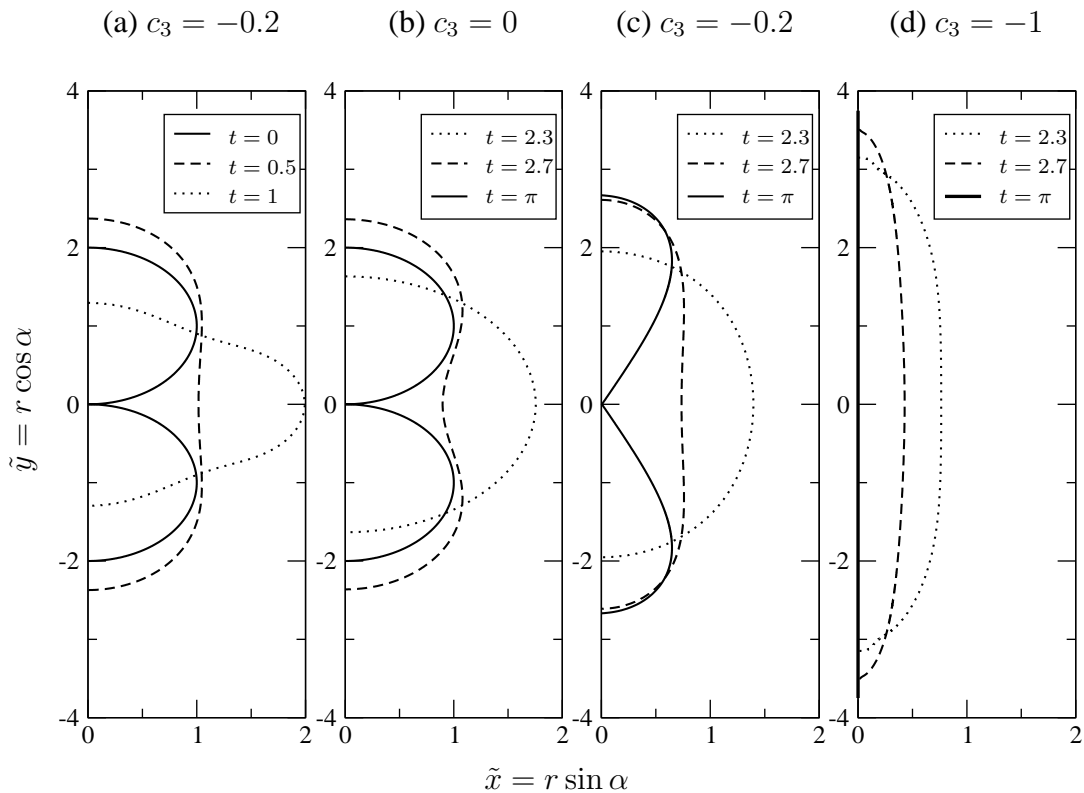}
 \caption{Embeddings of surfaces $t=\textrm{constant}$, $\rho_1-\rho_2=0,\pi$ in Euclidean space. The complete 2-surfaces are obtained by rotating the curves around the $\tilde y$-axis. Parameters: $R_0=1$, $c_1=2$.\label{fig:Einbettung}}
\end{figure}

A couple of examples for several parameter values is given in Fig.~\ref{fig:Einbettung}, where the solution $r=r(\alpha)$ is plotted in Cartesian coordinates $\tilde x$, $\tilde y$ in the form $\tilde x(\alpha)=r(\alpha)\sin\alpha$, $\tilde y(\alpha)=r(\alpha)\cos(\alpha)$. The resulting curves represent cross sections $\varphi=\textrm{constant}$ of the cylindrically symmetric embedded surface.

Panel (a) of Fig.~\ref{fig:Einbettung} shows the behaviour of the embedded surfaces in the limit $t\to 0$. For $t=1$ we obtain a 2-surface of spherical topology as expected. For smaller $t$, the ``equatorial'' circumference at $\tilde y=0$ decreases and finally reaches $0$ for $t=0$. Therefore, we interestingly observe that the embedding for $t=0$ corresponds to \emph{two} spheres instead of only one spherical surface. Indeed, \eqref{eq:emb} can be solved exactly for $t=0$ and leads to $\theta=\alpha$, $r=\sqrt{R_0 c_1}=\textrm{constant}$ independently of the value of $c_3$. Hence each of the components $\Sigma_1$ and $\Sigma_2$ corresponds to an entire sphere of radius $\sqrt{R_0 c_1}$ instead of only a hemisphere (as for $t>0$).

The situation near $t=\pi$ for $c_3=0$ (which is the ``Taub case'', see Sec.~\ref{sec:Taub}) is shown in panel (b). We see that the qualitative behaviour near $t=\pi$ for $c_3=0$ is the same as the behaviour near $t=0$ for arbitrary $c_3$: we have surfaces of spherical topology which narrow down at the equator in the limit and finally divide into two spheres. Equations \eqref{eq:emb} can also be solved exactly for $t=\pi$ and $c_3=0$. The solution is $\theta=\alpha$, $r=\sqrt{R_0/c_1}=\textrm{constant}$.

The behaviour near $t=\pi$ is slightly different for $c_3\neq 0$, see Fig.~\ref{fig:Einbettung}c. In this case we again obtain spherical surfaces that approach a surface with two spherical components. However, the limiting surface is badly behaved at $\theta=0$ (corresponding to $\tilde x=\tilde y=0$), where a \emph{conical singularity} is present, i.e.\ the curves are not orthogonal to the $\tilde y$-axis at this point. 

A special case is the ``singular case'' $c_3=-1$. As illustrated in Fig.~\ref{fig:Einbettung}d, the 2-surfaces contract to an interval on the $\tilde y$-axis for $t\to\pi$. This also follows from \eqref{eq:metrich}, because the coefficient of $\dd\varphi^2$ tends to $0$ for $t\to\pi$ such that the two-metric degenerates to $h=\ee^M\dd\theta^2$. Obviously, this is the metric of a one-dimensional line. The reason is that the tangent vector $\partial_{\rho_1}+\partial_{\rho_2}$ on $\Sigma$ becomes a null vector for $t=\pi$, i.e.\ one of the two directions within $\Sigma$ becomes lightlike and does not contribute to the distance anymore. Moreover, we observe the expected singular behaviour of the solution at $t=\pi$, $\theta=0$ (corresponding to the ``north'' and ``south poles'' of the embedded figures), where the curvature of the embedded surfaces diverges. Indeed, the Gaussian curvature at the poles turns out to be $c_1(1-3c_3)/[R_0(1+c_3)^3]$ and diverges for $c_3\to -1$.

\subsection{The singular cases}
\label{sec:singularcases}

In our previous discussion we have mostly assumed that $c_3\neq\pm1$ holds, i.e.\ we have excluded the singular cases. But in the following we will have a closer look at them. 

As a consequence of the symmetry \eqref{eq:sym}, it is sufficient to discuss only the solutions with $c_3=1$. The models with $c_3=-1$ will differ from these only by a reflection at $\theta=\pi/2$, i.e.\ an interchange of the two axes, and a $\rho_2$-reflection.

We have seen in Sec.~\ref{sec:regularity} that the Kretschmann scalar $K$ diverges at point $C$ ($\theta=0$, $t=\pi$, or, equivalently, $x=1$, $y=-1$) for $c_3=1$. It is interesting to study the behaviour of $K$ in a vicinity of the singularity in more detail. To this end, starting from the $x$-$y$-coordinates, we introduce polar coordinates $(r,\phi)$ centered at the point $C$,
\begin{equation}\label{eq:polar}
 x=1-r\cos\phi,\quad y=-1+r\sin\phi,\quad r\ge0, \quad\phi\in\Big[0,\frac{\pi}{2}\Big].
\end{equation}
In terms of these coordinates, the Kretschmann scalar becomes a rational function of $r$, $\sin\phi$ and $\cos\phi$, i.e.\ it has a simple structure even though the explicit expression is rather lengthy. The leading order behaviour close to the singularity at $r=0$ is given by
\begin{equation}\label{eq:exp}
 K = \frac{g(\phi)}{r^6}+\mathcal O(r^{-5})
\end{equation}
with
\begin{equation}
 g(\phi)=\frac{768c_1^6(c_1^2-4)(1+T^2)^3}{R_0^2(c_1^2+4)^2(4+c_1^2T^2)^6}p_1(c_1T)p_2(c_1T),\quad T=\tan\phi,
\end{equation}
where
\begin{equation}\fl
 p_{1/2}(x)=x^3-6\alpha_{1/2}x^2-12x+8\alpha_{1/2},\quad
 \alpha_1=\frac{c_1-2}{c_1+2},\quad \alpha_2=-\frac{c_1+2}{c_1-2}.
\end{equation}
Note that, as a consequence of the rational structure of the full expression for $K$, the expansion \eqref{eq:exp} is not only valid for constant $\phi$, but actually holds uniformly in $\phi$. Hence, if we approach the singularity along an arbitrary curve $r(s)$, $\phi(s)$, where $s$ is some curve parameter, then the divergent behaviour of $K$ is determined by the behaviour of $g(\phi(s))$. Of particular importance are the zeros of $g(\phi)$, which are identical with the zeros of the polynomial $p_1(x)p_2(x)$. $p_1$ and $p_2$ are polynomials of third degree, and they turn out to always have three real zeros --- if we exclude the special case $c_1=2$ for a moment. Moreover, the zeros of $p_1(x)$ and $p_2(x)$ are distinct, hence the product $p_1(x)p_2(x)$ has six distinct real zeros. However, since $c_1>0$ and $T=\tan\phi\ge0$ for $\phi\in[0,\pi/2]$, the argument $c_1T$ must be non-negative, i.e.\ we are only interested in non-negative zeros. Since precisely three of the six zeros turn out to be positive, we 
see that there are always three directions $\phi_1$, $\phi_2$, $\phi_3$, along which the leading order term $\propto r^{-6}$ of $K$ vanishes, such that $K$ may then diverge at most proportional to $r^{-5}$. Moreover, at these zeros, the sign of $p_1(c_1T)p_2(c_1T)$ changes, i.e.\ there are both regions in which $K$  diverges to $+\infty$  and regions where it diverges to $-\infty$.
 
So far we have assumed $c_1\neq 2$. Now we look at the special case $c_1=2$, in which $g(\phi)$ simplifies to
\begin{equation}
 g(\phi)=\frac{192}{R_0^2(1+T^2)^3}T(T^2-3)(3T^2-1).
\end{equation}
The non-negative zeros are then located at $T=0, 1/\sqrt{3}, \sqrt{3}, \infty$, corresponding to $\phi=0,\pi/6,\pi/3, \pi/2$. Hence, for $c_1=2$, the function $g(\phi)$ has four non-negative zeros.

The discussion so far shows that $K$ can diverge to $\pm\infty$, depending on the curve along which the singularity is approached. But could there even be curves along which $K$ remains finite? Such a curve would necessarily have to approach the singularity asymptotically along one of the directions given by the zeros of $g(\phi)$, since evidently the vanishing of the leading divergent term $\propto r^{-6}$ of $K$ is a necessary condition for $K$ to remain finite. And the behaviour of such a curve would need to be sufficiently ``fine-tuned'' near the singularity to achieve that also the other divergent terms $\propto r^{-5}$, $\propto r^{-4},\dots,\propto r^{-1}$ in $K$ vanish. Remarkably, this turns out to be possible, and we will illustrate this in the special case $c_1=2$, where the relevant formulae become simpler. By way of example, we give some curves with the desired properties in the form $x=x(y)$ or $y=y(x)$.

\begin{figure}\centering
 \includegraphics[scale=.9]{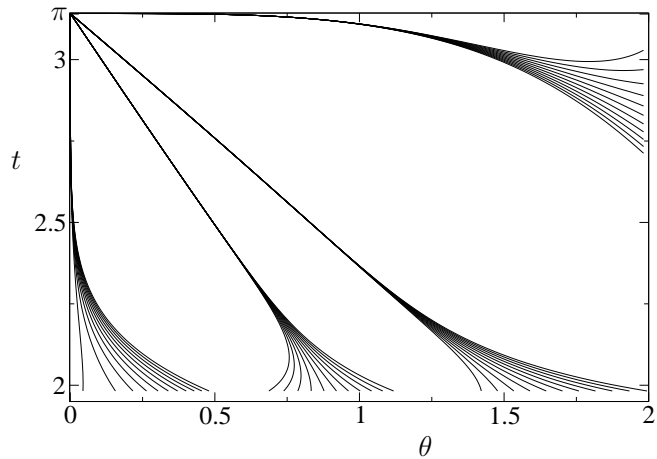}
 \caption{Illustration of the four families of curves $\gamma_1,\dots,\gamma_4$, cf.~\eqref{eq:curve1}-\eqref{eq:curve4}, along which the Kretschmann scalar in the case $c_1=2$ approaches a finite limit at the singularity. \label{fig:Curves}}
\end{figure}

The following four families of curves $\gamma_1,\dots,\gamma_4$, which depend on an additional constant parameter $z\in\R$, indeed all lead to a bounded Kretschmann scalar. The limit of $K$ as $x\to1$, $y\to-1$, which depends on $z$, is also indicated:

\begin{eqnarray}\fl\label{eq:curve1}
\gamma_1:\quad && x = 1-\frac{1}{96}(y+1)^4-\frac{1}{48}(y+1)^5-\frac{13}{768}(y+1)^6+z(y+1)^7,\\ \fl\nonumber
               &&  K \to -\frac{3}{4R_0^2}(768z+5), \\ \fl
 \gamma_2:\quad && x=1-\frac{\sqrt{3}}{3}(y+1)+\frac{1-\sqrt{3}}{6}(y+1)^2+\frac{2-\sqrt{3}}{12}(y+1)^3\\ \fl\nonumber
                && \qquad +\left(\frac{5}{36}-\frac{53}{648}\sqrt{3}\right)(y+1)^4+\left(\frac{175}{1296}-\frac{19}{162}\sqrt{3}\right)(1+y)^5
                   \\\fl\nonumber
               && \qquad +\left(\frac{973}{5184}-\frac{781}{5184}\sqrt{3}\right)(y+1)^6
                   +z(y+1)^7,\\ \fl\nonumber
               && K\to \frac{1}{768R_0^2}(139968z-53496+19261\sqrt{3}),\\ \fl
 \gamma_3:\quad && x=1-\sqrt{3}(y+1)+\frac{3-\sqrt{3}}{2}(y+1)^2+\frac{6-5\sqrt{3}}{4}(y+1)^3\\ \fl \nonumber
           &&\qquad +\left(\frac{49}{12}-\frac{15}{8}\sqrt{3}\right)(y+1)^4
            +\left(\frac{347}{48}-\frac{125}{24}\sqrt{3}\right)(y+1)^5\\ \fl\nonumber
           &&\qquad +\left(\frac{3155}{192}-\frac{2233}{192}\sqrt{3}\right)(y+1)^6+z(y+1)^7,\\ \fl\nonumber
        && K\to-\frac{1}{256R_0^2}(576z-25500+11695\sqrt{3}),\\ \fl
        \label{eq:curve4}
 \gamma_4:\quad && y=-1+\frac{1}{96}(x-1)^4-\frac{1}{192}(x-1)^5-\frac{5}{768}(x-1)^6
            +z(x-1)^7,\\ \fl\nonumber
        && K\to-\frac{9}{8R_0^2}(512z-5). 
\end{eqnarray}
Some of the curves in each family are illustrated in Fig.~\ref{fig:Curves}. One can clearly see that curves of the same family are almost indistinguishable close to the singularity, since they have to approach this point in a well-defined way to guarantee regularity of the Kretschmann scalar. Note that the directions, along which the four families approach the singularity, correspond to the four non-negative zeros of $g(\phi)$ in this case.
Since the limit of the Kretschmann scalar is a linear function of $z$ in all four cases, the limit can be any real number. 

We conclude that we can approach the singularity at point $C$ either along curves such that $K\to\pm\infty$, or along curves such that $K$ has any prescribed finite limit. In other words, we observe a \emph{directional behaviour} of the Kretschmann scalar. This is similar to the behaviour of the Kretschmann scalar in the Curzon solution \cite{Curzon1925}, where it turned out that the singularity contains some ``hidden structure'', and it is actually possible to extend the solution beyond that singularity. The original construction of the extended Curzon spacetime by Scott and Szekeres can be found in \cite{ScottSzekeres1986a,ScottSzekeres1986b}, and for a detailed overview we refer to \cite{Whale2014}. The Curzon singularity was classified later on as a so-called \emph{directional singularity}. Roughly speaking, this means that it is possible to approach the singularity either along curves such that the curvature becomes singular (e.g.\ the Kretschmann scalar diverges) or along curves at which 
everything remains regular. Moreover, it is possible to extend the spacetime through the singularity and to reach further regular regions. For a precise definition and explanation of directional singularities from the  point of view of abstract boundary constructions, we refer to \cite{ScottSzekeres1994} and to \cite{Ashley2002,Whale2010}.

The directional behaviour described above for our smooth Gowdy-symmetric generalized Taub-NUT solution might lead to the conjecture that the singularity for $c_3=1$ is also a directional singularity. However, for this it is not enough that the Kretschmann scalar remains finite along some curves. Instead, the entire geometry must remain regular. In particular, there must be curves approaching the directional singularity along which \emph{every} curvature invariant is bounded.  Hence, if we could only find one invariant that diverges, even though the Kretschmann scalar remains finite, we would not have a directional singularity. Interestingly --- or unfortunately, if one likes directional singularities --- such a quantity can indeed be provided, namely the invariant
\begin{equation}
 J:=R^{ab}{}_{cd} R^{cd}{}_{ef}R^{ef}{}_{ab},
\end{equation}
which is cubic in the Riemann tensor, in contrast to $K$, which is quadratic. The explicit calculation shows that
\begin{equation}
 J=\frac{\tilde P(x,y)}{\ee^{9M(x,y)}},
\end{equation}
where $\tilde P$ is another (very lengthy) polynomial in $x$ and $y$ (of 24th degree in $x$ and 36th degree in $y$). In terms of the polar coordinates \eqref{eq:polar}, $J$ becomes
\begin{equation}
 J=\frac{\tilde g(\phi)}{r^9}+\mathcal O(r^{-8}),
\end{equation}
with
\begin{eqnarray}
 \tilde g(\phi) &=& -\frac{6144 c_1^9(1+T^2)^{9/2}}{R_0^3(c_1^2+4)^3(4+c_1^2T^2)^9}\tilde p_1(T)\tilde p_2(T),\\
 \tilde p_1(T) &=& c_1^4T^3+12c_1^2T^2-12c_1^2T-16,\\
 \tilde p_2(T) &=& c_1^6(c_1^2-12)T^6+96c_1^6T^5-12c_1^4(11c_1^2-36)T^4-1280c_1^4T^3\\
               &&+48c_1^2(9c_1^2-44)T^2+1536c_1^2T-192c_1^2+256.\nonumber
\end{eqnarray}
Again, the expansion holds uniformly in $\phi$. The zeros of $\tilde g(\phi)$ correspond to directions in which $J$ diverges slower, or, potentially, remains finite --- similarly to the above mentioned properties of $K$ in relation to the zeros of $g(\phi)$. Therefore, a necessary condition for the existence of a curve along which both $K$ \emph{and} $J$ remain finite is a simultaneous zero of $\tilde g(\phi)$ and $g(\phi)$. However, as appropriate combinations of the polynomials $p_1$ or $p_2$ with $\tilde p_1$ or $\tilde p_2$ reveal, there are no values of the parameter $c_1$ for which such simultaneous zeros exist. Hence, $J$ necessarily diverges along all curves on which $K$ remains finite and vice versa.

We conclude that, even though the Kretschmann scalar exhibits some directional behaviour, the singularity is actually not a directional singularity in the strict sense. In particular, there cannot be any reasonable extension of the spacetime through the singularity. This provides an interesting example of a solution for which the Kretschmann scalar does not contain all information about the singular behaviour, but where, in addition, other invariants have to be studied.

\subsection{Beyond the Cauchy horizons\label{sec:extensions}}
{In all of our previous discussion we have considered the situation between the two boundaries $t=0$ and $t=\pi$, where, unless in the singular cases, smooth Cauchy horizons are located}. However, since it is a general property of smooth Gowdy-symmetric generalized Taub-NUT solutions that they can be extended through the horizons \cite{beyer11}, it might also be interesting to study some properties of \emph{extensions} of our exact solution.

For a given spacetime $(M,g)$, another spacetime $(\hat M,\hat g)$ is called an extension of $(M,g)$, if there exists an isometric embedding $\Lambda:M\to\hat M$ and if $\hat M$ is ``larger'' than $M$ in the sense that $\Lambda(M)\subsetneqq \hat M$. We refer to the article by Chru\'sciel and Isenberg \cite{chrusciel93} for detailed definitions and discussions of spacetime extensions. In particular, extensions of the Taub solutions have been investigated in \cite{chrusciel93}. There are two ``standard'' past extensions and two ``standard'' future extensions of the Taub solutions. Chru\'sciel and Isenberg have shown that the two standard past extensions are equivalent (i.e.\ related via an isometry), and also the two standard future extensions are equivalent. Combining the two future and the two past extensions, one obtains four spacetimes that include both types of extensions. Interestingly, these four extensions can be divided into two groups. Both groups contain two equivalent extensions. However, each 
extension in the first group is \emph{not} equivalent to either extension in the second group. Hence there are \emph{inequivalent} extensions of the Taub spacetimes. In the following we will see that our exact solution has similar properties.

The starting point for the construction of extensions of arbitrary smooth Gowdy-symmetric generalized Taub-NUT solutions in \cite{beyer11} was the observation that the representation of the metric \eqref{eq:metric} in terms of our coordinates is singular at $t=0$ and $t=\pi$, where $\det(g)=0$ holds, see \eqref{eq:det}. However, it is possible to introduce new coordinates $(t',\theta,\rho_1',\rho_2')$ with respect to which we can extend the solution in a regular way through the Cauchy horizons. An extension is then obtained by extending the domain of $t'$ and keeping the same form of the metric also for $t'$-values that correspond to points beyond the Cauchy horizons. To this end, appropriate smooth extensions of the metric potentials $M$, $u$ and $Q$ also have to be chosen. The required isometry $\Lambda$ is then just given by the identity map $(t',\theta,\rho_1',\rho_2')\in M\mapsto (t',\theta,\rho_1',\rho_2')\in\hat M$.

Here we follow the same idea and construct extensions of our solution by first introducing new coordinates. However, instead of the new time coordinate $t'$ from \cite{beyer11}, we can also use $y=\cos t$. This is particularly useful since all metric potentials are already given as functions of $y$ (and $x=\cos\theta$, for which reason we will also use $x$ as new coordinate, even though this would not be necessary for guaranteeing regularity at the Cauchy horizons). In addition, we perform transformations of the coordinates $\rho_1$ and $\rho_2$, which will be given shortly. The potentials $M$, $u$ and $Q$ are then extended from the domain $y\in(-1,1)$ to $y\in\R$ by choosing their analytic continuations (which is possible in our case, since our solution is not only smooth but even analytic). In other words, we use the same formulae \eqref{eq:solnew1}, \eqref{eq:solnew2} also for $|y|\ge1$.
In the following, we separately discuss future extensions, past extensions, and combinations of both.

We start by extending the solution through the past Cauchy horizon at $t=0$ ($y=1$). For that purpose, we introduce new coordinates $(x,y,\rho_1',\rho_2')$ via
\begin{equation}\label{eq:tran1}
 x=\cos\theta,\quad
 y=\cos t,\quad
 \rho_1=\rho_1'+\kappa\ln(1-y),\quad
 \rho_2=\rho_2',
\end{equation}
where $\kappa=\mathrm{constant}$. In terms of these coordinates, the metric becomes
\pagebreak
\begin{eqnarray}\fl
 g &=& \frac{\ee^M}{1-x^2}\,\dd x^2
      +\frac{R_0\kappa^2(1+y)^2\ee^u-\ee^M}{1-y^2}\,\dd y^2
     \\ \fl\nonumber
   &&
      +R_0\Big[-2\kappa(1+y)\ee^u(\dd\rho_1'+Q\dd\rho_2')\dd y
      +(1-y^2)\ee^u(\dd\rho_1'+Q\dd\rho_2')^2+(1-x^2)\ee^{-u}\dd\rho_2'^2\Big].
\end{eqnarray}
The apparently singular component $g_{yy}$ remains regular at $y= 1$ if we choose
\begin{equation}\label{eq:kap}
 \kappa=\pm\sqrt{\lim\limits_{y\to 1}\frac{\ee^{M-u}}{4R_0}}=\pm\frac{c_1}{2}.
\end{equation}
Note that this is only possible because $\lim_{y\to 1}\ee^{M-u}$ does not depend on $x$, so that $\kappa$ is indeed a constant. However, this is not a coincidence for our particular solution, but holds in general for all smooth Gowdy-symmetric generalized Taub-NUT solutions as a consequence of the Einstein equations. 

The above coordinate transformation removes the coordinate singularity at the past Cauchy horizon. Consequently, we will use the transformed version of the metric also for $y\ge1$, i.e.\ in the region beyond the past Cauchy horizon. This provides us with the required extension. Note that the two possible sign choices for $\kappa$ correspond to two different past extensions. Adopting the notation from \cite{chrusciel93}, we denote these as $(M^{\downarrow\pm},g^{\downarrow\pm})$, where `$\pm$' specifies the sign of $\kappa$.

Using the explicit solution, it is easy to show that the metric coefficients have no singularities in the extended region (with exception of the usual axes singularities at $x=\pm 1$, which could be removed by another coordinate transformation), provided $\ee^M\neq 0$ holds. Moreover, the determinant of the metric is now $\det(g)=-R_0^2\ee^{2M}$, i.e.\ the metric is invertible wherever  $\ee^M\neq 0$ holds. Therefore, the question of regularity of our extensions reduces to a discussion of zeros of $\ee^M$. We have seen earlier that, in the regular cases with $c_3\neq\pm1$, $\ee^M$ has no zeros inside the Gowdy square. However, there might be zeros in the extension $y>1$. And indeed, we show in \ref{App1} that there is precisely \emph{one} zero in this region. Since we have already seen that the Kretschmann scalar diverges at zeros of $\ee^M$, cf.\ \eqref{eq:Kret}, these zeros do not correspond to mere coordinate singularities, but to physical curvature singularities. Hence we conclude that \emph{there is 
always one singularity in each of our two past extensions}. This singularity can be represented as a point in an $x$-$y$ diagram, but due to the additional degrees of freedom $\rho_1'$ and $\rho_2'$, it actually has the topology of a 2-torus. {(An exception are singularities on the axes, which appear exclusively in the singular cases $c_3=\pm1$. These have the topology of a circle.)}

In a next step, we consider a future extension of our solution. To this end, we perform a slightly different coordinate transformation,
\begin{equation}\label{eq:tran2}\fl
 x=\cos\theta,\quad
 y=\cos t,\quad
 \rho_1=\rho_1'+\kappa_1\ln(1+y),\quad
 \rho_2=\rho_2'+\kappa_2\ln(1+y)
\end{equation}
with constants $\kappa_1$ and $\kappa_2$. Similarly to the above discussed past extension, this transformation removes the coordinate singularity at the future Cauchy horizon \mbox{($y=-1$)}, provided we choose
\begin{equation}\label{eq:kap12}\fl
 \kappa_2=\pm\sqrt{\lim\limits_{y\to -1}\frac{(1-y^2)\ee^{M+u}}{4R_0(1-x^2)}}
         = \pm\frac{c_3}{c_1},\quad
 \kappa_1=-\kappa_2\lim\limits_{y\to -1}Q=-\frac{c_3^2+1}{2c_3}\kappa_2.
\end{equation}
Again we have the freedom to choose a sign, which gives rise to two different future extensions. We denote these as $(M^{\uparrow\pm},g^{\uparrow\pm})$, where `$\pm$' indicates the sign of $\kappa_2$. 

An investigation of the transformed metric shows that its regularity is again equivalent to $\ee^M\neq 0$. We show in \ref{App1} that $\ee^M$ has either one or two zeros in the future extension, depending on the value of $c_3$. Hence, \emph{there is at least one curvature singularity in each of our two future extensions}.

As mentioned above, in the case of the Taub solution (i.e.\ the special case $c_3=0$ of our solution), the two standard future/past extensions are equivalent. We can easily show that this is also true for our future/past extensions with general $c_3\in\R$. The extensions $(M^{\downarrow+},g^{\downarrow+})$ and $(M^{\downarrow-},g^{\downarrow-})$ are related via the isometry $(\rho_1',\rho_2')\mapsto(-\rho_1',-\rho_2')$ and therefore equivalent. Similarly, the extensions $(M^{\uparrow+},g^{\uparrow+})$ and $(M^{\uparrow-},g^{\uparrow-})$ are equivalent, which follows from the same isometry.

Finally, we look at simultaneous past and future extensions. These could be constructed by pasting together one of our past extensions with one of the future extensions, which leads to four different spacetimes. Each of these would be described in terms of two coordinate patches, namely one for the past region and one for the future region. However, it is even possible to obtain past and future extensions for which  
 a single coordinate patch is sufficient. To this end, we start again from our original, not yet extended solution and perform the coordinate transformation
\begin{equation}\fl\label{eq:tran3}
 x=\cos\theta,\
 y=\cos t,\
 \rho_1=\rho_1'+\kappa\ln(1-y)+\kappa_1\ln(1+y),\
 \rho_2=\rho_2'+\kappa_2\ln(1+y),
\end{equation}
which essentially combines the earlier transformations \eqref{eq:tran1} and \eqref{eq:tran2}. With the same choices for the constants $\kappa$, $\kappa_1$ and $\kappa_2$ as before we arrive at an extended spacetime that is regular wherever $\ee^M\neq 0$. We denote these extensions as $(M^{ab},g^{ab})$, where $a=+,-$ determines the sign of $\kappa$ and $b=+,-$ the sign of $\kappa_2$. Note that both extensions $(M^{+b},g^{+b})$, $b=+,-$, when restricted to $y>-1$, are basically the same as our earlier past extension $(M^{\downarrow +},g^{\downarrow +})$, since they only differ by a \emph{regular} coordinate transformation $\rho_1'\mapsto\rho_1'+\kappa_1\ln(1+y)$, $\rho_2'\mapsto\rho_2'+\kappa_2\ln(1+y)$. In the same way, the two extensions $(M^{-b},g^{-b})$, restricted to $y>-1$, both correspond to $(M^{\downarrow -},g^{\downarrow -})$. Similar statements apply to the restriction of $(M^{a\pm},g^{a\pm})$ to $y<1$ and our two future extensions. 

The above mentioned remarkable result in \cite{chrusciel93} for extensions of the Taub solution was that
\begin{enumerate}
 \item $(M^{++},g^{++})$ is equivalent to $(M^{--},g^{--})$,
 \item $(M^{+-},g^{+-})$ is equivalent to $(M^{-+},g^{-+})$,
 \item there are no isometries between the other pairs of extensions.
\end{enumerate}
The statement (iii) might be particularly surprising, given that the ingredients of the global extensions, namely the two future extensions the two past extensions, are equivalent, respectively. 

It is easily possible to generalize (i) and (ii) to our solution. This follows immediately from the isometry $(\rho_1',\rho_2')\mapsto(-\rho_1',-\rho_2')$.
The interesting question now is whether (iii) also applies in our situation. The proof of (iii) in the Taub case made essential use of properties of null geodesics and their extendibility through the Cauchy horizons. Since the Taub solution has four Killing vectors, there are enough conservation laws to determine all geodesics up to quadrature, see \cite{MisnerTaub1969}. In our case, however, there are ``only'' two Killing vectors, which makes the calculation of geodesics more complicated. Hence we do not aim for a rigorous proof of (iii) for our solution. However, the behaviour of those special null geodesics that are restricted to the axes $\theta=0,\pi$ is very similar to the geodesics of the Taub solution. This is discussed in \ref{App2}. And it might well turn out to be sufficient to study the extendibility of \emph{axis} geodesics to prove (iii)%
\footnote{Following the idea of Chru\'sciel and Isenberg's proof of (iii) in the Taub case, we would need to show that a hypothetical isometry between, say, $M^{++}$ and $M^{+-}$ necessarily maps an axis geodesic to an axis geodesic. This could possibly be shown using the following observation. The boundaries $x=\pm 1$, $y=\pm 1$ of the Gowdy square can be characterized in terms of the Killing vectors $\xi$, $\eta$ (where $\xi=\partial_{\rho_1}$, $\eta=\partial_{\rho_2}$ in our coordinates) as zeros of $W:=g(\xi,\xi) g(\eta,\eta)-g(\xi,\eta)^2$. Note that $W$ is a scalar and therefore invariant under coordinate transformations. Moreover, also a change of the Killing basis leaves the zeros of $W$ invariant, since $W$ is then only multiplied by a positive factor. Hence the hypothetical isometry would map the coordinate set $x=\pm 1$, $y=\pm 1$ of $M^{++}$ to the same set in $M^{+-}$. However, a rigorous extension of Chru\'sciel and Isenberg's proof of (iii) to our situation is beyond the scope of 
this paper.}. 
Based on these observations, we conjecture that also (iii) generalizes to our solution.

Finally we note that, as in the case of the Taub solutions, the extensions contain \emph{closed causal curves}, i.e.\ there are problems in terms of causality. As an example, consider the curve
\begin{equation}
 x(s)=0,\quad
 y(s)=1+\frac{2c_1}{|c_3|},\quad
 \rho_1'(s)=s,\quad
 \rho_2'(s)=0,\quad s\in[0,4\pi]
\end{equation}
in our \emph{past} extensions. Due to the periodicity of the $\rho_1'$ and $\rho_2'$ coordinates, this curve is closed. Moreover, we have $g_{\rho_1'\rho_1'}=-\frac{4|c_3|R_0}{1+c_3^2}<0$ along the curve, i.e.\ the tangent vectors are indeed timelike. Similarly,
\begin{equation}\fl
 x(s)=0,\quad
 y(s)=y_0=\mbox{constant}\ll -1,\quad
 \rho_1'(s)=0,\quad
 \rho_2'(s)=s,\quad s\in[0,4\pi]
\end{equation}
is an example of a closed timelike curve in our \emph{future} extensions. (Because of $g_{\rho_2'\rho_2'}|_{x=0}=-\frac{R_0 c_3^2}{16c_1}y^4+\mathcal O(y^3)$, this curve is timelike for sufficiently negative $y_0$.)

\subsection{Leading-order behaviour and spikes}
\label{sec:spikes}

Another way of looking at the exact  solution found in this paper is to derive the ``leading-order behaviour'' and hence expansions at $t=0$ and $\pi$. The formulation of the vacuum equations as a singular initial value problem  in \cite{beyer11} (Theorem~3.1 there) and the expansions given in \Theoremref{Thm1} here with asymptotic data $S_{**}$ and $Q_*$ gives rise to this in the case of $t=0$; the expansion of the function $\omega$ is given by Proposition~3.2 in \cite{beyer11} in terms of an irrelevant constant $\omega_*$ and the data function $\omega_{**}$ related to $Q_*$ as
\[\omega_{**}(\theta)=\ee^{2S_{**}(\theta)}\frac{1-\partial_\theta Q_*(\theta)\sin\theta
  -2Q_*(\theta)\cos\theta}{4R_0}.\]
The variables $\lambda$, $\omega$ and $Q$ here are defined with respect to the choice $\xi_1=\partial_{\rho_1}$ and $\xi_2=\partial_{\rho_2}$; indeed, one of the assumptions which was made in \cite{beyer11}  is that $\partial_{\rho_1}$ is parallel to the generator of the past horizon.
For the explicit solution in this paper now, which is expressed with respect to the same choice $\xi_1$ and $\xi_2$, we shall indeed confirm these expansions at $t=0$ below.

Concerning expansions at  $t=\pi$, however, we expect a different behaviour of $\lambda$, $Q$ and $\omega$ defined with respect to the same choice of $\xi_1$ and $\xi_2$ since, in general, $\partial_{\rho_1}$ is not parallel to a generator of the horizon at $t=\pi$. 
Indeed, we find ``spiky features''. Expressing these quantities with respect to a different choice of $\xi_1$ and $\xi_2$, however, removes those, at least in the ``regular'' cases $c_3\not=\pm 1$, and we find analogous expansions as at $t=0$ (except for some minor differences due to different topological properties of the generators). This is consistent with the established idea \cite{RendallWeaver2001} that spikes which can be removed by a change of the Killing bases are \keyword{false} spikes.

In the ``singular cases'' $c_3=\pm 1$, the spiky features, which we identify below at $t=\pi$, cannot be removed by a change of the Killing basis and hence those are \keyword{true} spikes.

\subsubsection{Some background.}
 
A consequence of the condition $[\xi_1,\xi_2]=0$ for a general choice of $\xi_1$ and $\xi_2$ according to \Eqref{eq:transformedbasis}
is that both fields generate alternative coordinates $\phi_1$, $\phi_2$ on the symmetry orbits with $\xi_1=\partial_{\phi_1}$ and $\xi_2=\partial_{\phi_2}$. The metric $g$ can then locally be written in a very similar manner as before in \Eqref{eq:metric}:
\begin{equation*}
  g=\ee^M(-\dd t^2+\dd\theta^2)
  +\tilde\lambda (\dd{\phi_1}+\tilde{Q} \dd{\phi_2})^2+\frac{\tilde{R}^2}{\tilde\lambda} \dd{\phi_2}^2,
\end{equation*}
with 
\[\tilde R=\tilde R_0 \sin t\sin\theta,\]
for some $\tilde R_0>0$, where
\begin{equation}
  \label{eq:relationPQomegatheta}
 \partial_t\tilde\omega = -\tilde R^{-1}\tilde\lambda^2\,\partial_\theta \tilde Q,\quad
 \partial_\theta\tilde\omega = -\tilde R^{-1}\tilde\lambda^2\,\partial_t \tilde Q,
\end{equation}
are the quantities defined with respect to a general choice of $\xi_1$ and $\xi_2$ according to \Eqref{eq:transformedbasis}, while, as we agree from now on, the corresponding quantities with no tilde refer to the particular choice $\xi_1=\partial_{\rho_1}$ and $\xi_2=\partial_{\rho_2}$.
Using the transformation laws in \cite{beyer11}, we derive
\[\fl\,\,\tilde R_0=|ad-bc|R_0,\quad
\tilde\lambda=(a+b Q)^2 \lambda+b^2 R^2\lambda^{-1},\quad
\tilde Q=\frac{(a+b Q) (c+d Q) \lambda+bd R^2 \lambda^{-1}}{(a+b Q)^2 \lambda+b^2 R^2 \lambda^{-1}}.\]
This and \Eqref{eq:relationPQomegatheta} then allows to compute $\tilde\omega$ by a line integration.

We have seen that for any choice of Killing vector fields $\xi_1$ and $\xi_2$ (under the conditions $a\not=\pm b$), Geroch's reduction leads to a smooth quotient manifold $S$ with a smooth projection map $\pi$, which, locally in adapted coordinates, looks like
\[\pi: M\rightarrow S,\quad (t,\theta,\phi_1,\phi_2)\mapsto (t,\theta,\phi_2).\]
Moreover, Einstein's vacuum equations imply that the pair $(\tilde \lambda,\tilde \omega)$ is a wave map into the half-plane model of hyperbolic space. In the following, we consider the hyperbolic speed $\tilde s$ as defined in \Eqref{eq:hypspeed} with respect to the curves $t=2\arctan\ee^\tau$, and $\theta,\phi_2=\mathrm{constant}$ on $S$. 
Since $(\tilde\lambda,\tilde\omega)$ is a wave map with respect to the hyperbolic metric $(\dd\tilde\lambda^2+\dd\tilde\omega^2)/\tilde\lambda^2$, it follows that  the pair $(\tilde S,\tilde \omega)$ is a wave map into hyperbolic space with the metric $\dd \tilde S^2+\ee^{-2\tilde S}\dd\tilde \omega^2$ (this is just a change of coordinates on hyperbolic space).
Hence, the hyperbolic speed $\tilde s$ becomes 
\begin{equation}
\label{eq:ourspeed}
\tilde s(t,\theta)=\sin t\, \sqrt{\frac{[\partial_t \tilde \lambda(t,\theta)]^2+[\partial_t \tilde\omega(t,\theta)]^2}{\tilde\lambda^2(t,\theta)}}.
\end{equation}

Before we analyse the behaviour of all the quantities at $t=0$ and $\pi$, we notice that 
the explicit formulae in \Sectionref{sec:metricpotentials} allow us determine 
\[\lambda(x,y)=R_0 (1-y^2)\ee^{u(x,y)}\]
from \Eqref{eq:solu}; recall that $y=\cos t$ and $x=\cos\theta$. The function $Q(x,y)$ is given by \Eqref{eq:solQ}. 
By line integration of \Eqref{eq:relationPQomegatheta} and choosing the irrelevant constant appropriately, we find an explicit formula for the twist potential,
\begin{equation}
 \omega(x,y) = 16c_1R_0 \frac{(1-y)V}{U^2+V^2}
\end{equation}
with $U$ and $V$ as in \eqref{eq:defUV}.

\subsubsection{The behaviour at $t=0$.}
It is straightforward to determine expansions at $t=0$ and therefore confirm the results above. Consistent with our expectations, we find that the uniform expansions of \Theoremref{Thm1} hold at $t=0$ with 
\[\ee^{S_{**}(x)}=\frac{R_0}{c_1}, \quad Q_*(x)=\frac 32 c_3,\]
i.e.,
\[\omega_{**}(x)=\frac{{R_0} (1-3 c_3 x )}{4 c_1^2}.\]
The hyperbolic speed $s$ in \Eqref{eq:ourspeed} converges uniformly to the value $2$ at $t=0$.

\subsubsection{The behaviour at $t=\pi$ for $c_3\not=\pm1$.}
Let us stick with the choice $\xi_1=\partial_{\rho_1}$ and $\xi_2=\partial_{\rho_2}$ and determine the limit of the relevant functions at $t=\pi$ (i.e., $y\rightarrow-1$) first. We find that  
\[\lim_{y\rightarrow-1} \lambda(x,y)
=\frac{256 c_1 c_3^2 R_0 (1 - x^2)}{64 [c_1^2 (1 - c_3 x)^2 + c_3^4 (1 - x^2)^2]},\]
for every $x\in[-1,1]$.
Unless $c_3=\pm 1$, the convergence is uniform in space and the limit function is smooth. This is to be expected since the horizon at $t=\pi$ is smooth and $\partial_{\rho_1}$ (which, as mentioned earlier, can be defined without making reference to coordinates) extends as a smooth vector field to the future horizon. Since $\partial_{\rho_1}$ is not proportional to the generator at $t=\pi$ (except for $c_3=0$), the function $\lambda$ does not vanish (in contrast to the situation at $t=0$). 

Similarly, we find
\[\lim_{y\rightarrow-1}\omega(x,y)
=\frac{256 c_1^2 R_0 (1 - c_3 x)}{64 [c_1^2 (1 - c_3 x)^2 + c_3^4 (1 - x^2)^2]},\]
and the convergence to a smooth function (unless $c_3=\pm 1$) is uniform as before.

The limit of $Q$ for $c_3\not =0$ is
\[\lim_{y\rightarrow-1}Q(x,y)=\left\{
	\begin{array}{ll}
		(1 + c_3^2)/(2 c_3)  & \mbox{if } x\in (-1,1) \\
		1 & \mbox{if } x=1\\
                -1 & \mbox{if } x=-1.
	\end{array}
\right.
\]
Consequently, $Q$ cannot be extended as a continuous function to $t=\pi$. Only in the special case $c_3=0$ (the case of the Taub solution) we have $Q(x,y)=x$ so that $\lim_{y\rightarrow-1}Q(x,y)=x$ and therefore $Q$ extends smoothly. 

Despite the fact that $(\lambda,\omega)$ is a pair of smooth well-defined quantities through $t=\pi$, the hyperbolic speed defined with respect to it does not have continuous limit:
\begin{equation}
  \label{eq:limithyperbolicspeed}
  \lim_{y\rightarrow-1}s^2(x,y)=\left\{
	\begin{array}{ll}
		0  & \mbox{if } x\in (-1,1) \\
		4(1+4c_3^2/c_1^2) & \mbox{if } x=\pm 1.
	\end{array}
\right.
\end{equation}
This discontinuous behaviour of the hyperbolic speed at $t=\pi$ is interpreted as \keyword{spikes}. However, since the geometry is smooth at $t=\pi$ for $c_3\not=\pm1$, we claim that there exists another parametrization of the solution (i.e., another choice of $\xi_1$ and $\xi_2$ and hence another choice of wave map $(\tilde\lambda,\tilde\omega)$) for which this spiky behaviour disappears. Hence these are false spikes.

To this end, we choose 
\begin{equation}
  \label{eq:transfxi1}
  b=-2a c_3/(1+c_3^2)
\end{equation}
for the definition of $\xi_1$ and arbitrary $c$ and $d$ such that $ad-bc\not=0$ for the definition of $\xi_2$ in \Eqref{eq:transformedbasis}. Notice that this is compatible with the requirement $b\not=\pm a$ unless $c_3=\pm1$. Then we find that $\tilde\lambda$ is a smooth function through $t=\pi$ with the property
\[\lim_{y\rightarrow -1} \frac{\tilde\lambda(x,y)}{1+y}
=\frac{2 a^2 R_0 [c_1^2 (1 - c_3 x)^2 + c_3^4 (1 - x^2)^2]}{c_1 (1 + c_3^2)^2},\]
where the convergence is uniform in $x\in[-1,1]$.
Notice that this limit is in agreement with the expansions at $t=0$ before, namely, this limit function corresponds to the data function $\ee^{S_{**}}$ at $t=\pi$. The only new aspect is that in general the condition $S_{**}(0)=S_{**}(\pi)$, which was part of Theorem~\ref{Thm1},
 is violated. This is a consequence of the fact that the future horizon is not generated by $\partial_{\rho_1}$ but in fact by $\xi_1$ with the above choice of $b$.

Similarly, we obtain
\begin{eqnarray*}
\fl\,\,\lim_{y\rightarrow -1} \tilde Q(x,y)=
(1 + c_3^2) \Biggl(-2 b c_1^2 \Bigl(c_1^2 (-1 + c_3 x)^2 + 
       c_3^3 (-1 + x^2) (c_3 - 4 x + 3 c_3 x^2)\Bigr)\\ 
\fl\qquad+ d \Bigl(c_3^7 (-1 + x^2)^4 + 
       c_1^4 (-1 + c_3 x)^2 (-2 x + c_3 (-1 + x^2))\\ 
\fl\qquad\qquad+       c_1^2 c_3^2 (-1 + x^2) [4 x - 4 c_3^2 x (-2 + x^2) - 
          c_3 (3 + x^2) + c_3^3 (-1 - 5 x^2 + 2 x^4)]\Bigr)
\Biggr)\\
\Biggl/\Biggl({2 a [c_1^2 (-1 + c_3 x)^2 + c_3^4 (-1 + x^2)^2]^2}\Biggr),
\end{eqnarray*}
which is a smooth function and converges uniformly in $x$ unless $c_3=\pm1$. Next, we can use \Eqsref{eq:relationPQomegatheta} to determine the function $\tilde\omega$ by line integration. We refrain from giving the explicit expression here since this is very long.
Still, the fact that $\tilde w(x,y_0)$ must be a smooth function of $x$ for each $y_0\in (-1,1)$ together with the relation
\[\tilde\omega(x,y)=\tilde\omega(x,y_0)-\int_{y_0}^y \frac{\ee^{2\tilde S(y',x)}}{1-(y')^2}\tilde Q_x(y',x)\,\dd y',\]
and the above limits of $\ee^{\tilde S}$ and $\tilde Q$, implies that $\lim_{y\rightarrow -1}\tilde\omega (x,y)$ converges uniformly to a smooth function in $x$ unless $c_3=\pm 1$.

The limit $y\rightarrow-1$ of the hyperbolic speed $\tilde s$ is constant with value $2$ if $c_3\not=\pm 1$. 
This shows that the false spike behaviour completely disappears under the transformation \Eqref{eq:transfxi1}.

\subsubsection{The behaviour at $t=\pi$ for $c_3=\pm1$.}

Let us now proceed with the singular cases $c_3=\pm1$ and their behaviour at $t=\pi$.
The curvature singularity is located on one of the axes at
$t=\pi$, i.e., at $\theta=0$ for $c_3=1$ and at $\theta=\pi$ for $c_3=-1$. Let us for definiteness now restrict to the case $c_3=1$. In a first step let $\tilde\Sth$ be the set of points on $\Sth$ without the points corresponding to $\theta=0$. Then, we remove this axis from $M$, i.e., we define
\[\tilde M:=M\backslash \left((0,\pi)\times\tilde\Sth\right).\]
The restriction of the solution $g_{ab}$ to $\tilde M$ yields a smooth (but not globally hyperbolic) spacetime which satisfies Einstein's vacuum equations. We can show this can be extended smoothly through $t=\pi$. In fact $t=\pi$ corresponds to a smooth null hypersurface whose generator is proportional to $\partial_{\lambda_2}$. The field $\partial_{\lambda_2}$ never vanishes on $\tilde M$ since we have removed precisely those point on $M$ where it does. 
For the choice of Killing fields $\xi_1$ and $\xi_2$ as in \Eqref{eq:transfxi1} where now $\xi_1=\partial_{\lambda_2}$, the Geroch reduction is well-defined on $\tilde M$ and therefore yields a global smooth wave map structure as before. 
We find that the hyperbolic speed $\tilde s$ with respect to this choice converges pointwise on $\tilde M$ to the constant function $2$ at $t=\pi$ (i.e., $y=-1$). Geometrically, the singularity of the spacetime $M$ at  $t=\pi$ arises because $\partial_{\lambda_2}$, being parallel to the generator of the null hypersurface at $t=\pi$, describes smaller and smaller loops in the limit $\theta\rightarrow 0$, and hence, at $\theta=0$, the null generator does not have a well-defined direction.

What can we say about the solution at $t=\pi$ when we now consider the whole spacetime $M$ --- including the previously removed axis? The choice of $\xi_1$ and $\xi_2$ above, which is well-defined on $\tilde M$, yields a singular Geroch reduction at $\theta=0$. Nevertheless for our explicit solution, we can compute the limit $x\rightarrow 1$ (corresponding to $\theta=0$) of the hyperbolic speed $\tilde s$ at each $y\in (-1,1)$ (corresponding to $t\in(0,\pi)$). Surprisingly, it turns out the function which yields this limit of $\tilde s$ at every $y$ is smooth, and its limit $y\rightarrow -1$ is $4$. Hence, although the hyperbolic speed $\tilde s$ with respect to this choice of $\xi_1$ and $\xi_2$ extends nicely to the future horizon as a constant function with value $2$, it does not extend to a continuous function when the axis is taken into account. This discontinuity therefore appears precisely where the curvature is unbounded (\Sectionref{sec:singularcases}). It follows that in the 
singular 
cases $c_3=1$ (and similarly $c_3=-1$), the discontinuous behaviour at $t=\pi$ cannot be ``undone'' by a reparametrization of the Killing orbits and hence, the solution has a {true spike} at $t=\pi$ as expected.

\subsubsection{Comparison to the variables of St{\aa}hl.}

The work in \cite{Stahl02} by St{\aa}hl is the first attempt in the literature to formulate a singular initial value problem for Gowdy solutions with spatial $\SoXSt$-topology and $\Sth$-topology using Fuchsian techniques similar to the results obtained in \cite{KichenassamyRendall,rendall2000} for the case of spatial $\mathbb T^3$-topology. However, there are unexpected limitations as the results do not yield a family of solutions as large as expected from the $\mathbb T^3$-Gowdy case\footnote{St{\aa}hl also does not account for the constraint equations implied by the vacuum equations correctly. This problem is fixed, for a special class of solutions, in \cite{beyer11} which is also based on a different Fuchsian method \cite{AmesA,AmesB}.}.  St{\aa}hl conjectures that these are possibly related to the formation of spikes at the axes of symmetry under general conditions. In order to shed light on this let us study this question for our solution here. To this end, we must first relate the different sets of 
variables used here to these in \cite{Stahl02}.

St{\aa}hl chooses $\xi_1=\partial_{\lambda_1}$ and $\xi_2=\partial_{\lambda_2}$ (i.e., $a=b=c=1$, $d=-1$ in \Eqref{eq:transformedbasis}) throughout, for which, as described above, the Geroch reduction becomes singular at one of the axes. This, however, is not a problem, since St{\aa}hl focusses on the vicinity of the \emph{other} axis, namely the one at $\theta=0$. With respect to this basis of the algebra of Killing vectors, St{\aa}hl's quantity $X$ corresponds to our quantity $\tilde Q$, his quantity $Y$ corresponds to our $\tilde L:=\log\tilde\lambda-\tilde R$ and his $Z$ is the same as our $\tilde S:=\log\tilde\lambda$. In order to distinguish these quantities for this choice of Killing basis from the ones which we use above, we refer to them as $X$, $Y$ and $Z$ in the following.

Now there is an interesting relation between these sets of variables. Namely, it turns out that the pair $(\tilde L,\tilde Q)$ satisfies the wave map equations with the hyperbolic target metric $\dd \tilde L^2+\ee^{2\tilde L} \dd \tilde Q^2$ and the same source manifold as the pair $(\tilde S,\tilde \omega)$; one can check this easily by writing the wave map part of the field equations, namely \Eqsref{eq:Gerochevollambda} and \eref{eq:Gerochevolomega}, in terms of $\tilde L$ and $\tilde Q$ instead of $\tilde \lambda$ and $\tilde \omega$. However, since $\tilde R\sim\sin\theta$, the quantity $\tilde L$ is singular at each time $t$ at, at least, one of the axes. Hence $(\tilde L,\tilde Q)$ is a \textit{singular} wave map. This is different in the case of Gowdy solutions with $\mathbb T^3$-topology where $R=R_0 t$ and the main reason why the \keyword{Gowdy-to-Ernst transformation} does not work in the vicinity of the symmetry axes in the $\Sth$- and $\SoXSt$-Gowdy cases; for more details see \cite{
RendallWeaver2001}. 

Let us now go back to St{\aa}hl's parametrization of the solutions. He chooses to define a hyperbolic speed with respect to the singular pair $(\tilde L,\tilde Q)=(Y,X)$. In analogy with \Eqref{eq:ourspeed},  St{\aa}hl's  hyperbolic speed is (up to a sign)
\begin{equation*}
\nu(t,\theta)=\sin t \sqrt{[\partial_t Y(t,\theta)]^2+\ee^{2Y(t,\theta)}[\partial_t X(t,\theta)]^2}.
\end{equation*}

For our family of explicit solutions here, it turns out that the limit of the hyperbolic speed $\nu$ at $t=\pi$ is uniformly $1$ at $t=\pi$ if $c_3\not=\pm1$. In particular, this quantity can be unexpectedly extended continuously to the, in this parametrization, ``singular'' axis $\theta=\pi$. With respect to St{\aa}hl's variables the hyperbolic speed is therefore well behaved without signs of false spikes in contrast to our regular wave map parametrization above; cf.\  \Eqref{eq:limithyperbolicspeed}. If $c_3=1$, however, the limit of $\nu$ at $t=\pi$ is discontinuous at $\theta=0$ where the solution becomes singular: $\nu$ converges to the value $1$ everywhere except for $\theta=0$ (in particular, also in the same way as above, at $\theta=\pi$) and to the value $3$ along the axis $\theta=0$. In the case $c_3=-1$, the same discontinuity occurs at $\theta=\pi$. This is a further hint that these discontinuities must be considered as true spikes.  We notice that the limit values $1$ and $3$ of $\nu$, which 
we have found to occur for our solutions, are in consistency with St{\aa}hl's argument about the behaviour of general solutions at the axes in \cite{Stahl02}.

\section{Discussion\label{sec:discussion}}

We have derived an exact solution to Einstein's vacuum equations, which is a particular smooth Gowdy-symmetric generalized Taub-NUT solution. This was done by solving an initial value problem for the Ernst equation with ``Sibgatullin's integral method''. Our solution depends on three parameters $R_0>0$, $c_1>0$, $c_3\in\R$. For $c_3=0$, we arrive at the spatially homogeneous Taub spacetimes as a special case. Otherwise, we obtain spatially \emph{inhomogeneous} cosmological models. 

{We have shown that the solution is regular in the maximal globally hyperbolic region $0< t< \pi$. Moreover, the solution can be extended through $t=0$ and $t=\pi$ and has smooth Cauchy horizons at these surfaces.} Only in the ``singular cases'' $c_3=\pm 1$ are there scalar curvature singularities at the points $t=\pi$, $\theta=0$ or $t=\pi$, $\theta=\pi$. In these cases, the Kretschmann scalar $K$ shows a directional behaviour: $K$ diverges to $+\infty$ or $-\infty$ or to any real number, depending on the curve along which the singular point is approached. However, even if $K$ remains bounded, there are other scalars that diverge. Consequently, the singularities are not directional singularities in the sense used in abstract boundary constructions.

Furthermore, we have explicitly constructed several extensions of our solution. In particular, we have argued that it is likely that some of these extensions are not isometric, i.e.\ our solution seems to have inequivalent extensions, similar to the Taub spacetimes. Interestingly, for $c_3\neq 0$ all of our extensions contain singularities. We point out that there might be other extensions that are not isometric to the discussed ones. Moreover, among these there might be extensions that do not have singularities. However, we doubt that.

Moreover, this exact solution is an interesting example of an $\Sth$-Gowdy solution with spikes, both false and true spikes. In future research, this should therefore help to untangle the so far poorly understood relationship between the expected presence of spikes in generic situations and the behaviour at the axes of symmetry.

\section*{Acknowledgments}
We would like to thank Ben Whale, Gerrard Liddell {and J\"org Frauendiener} for valuable discussions and Gerrard Liddell for commenting on the manuscript. This work was supported by the Marsden Fund Council from Government funding, administered by the Royal Society of New Zealand.

\begin{appendix}

\section{Zeros of  \texorpdfstring{$\ee^M$}{exp(M)}\label{App1}}
The zeros of the potential $\ee^M$ determine the position of curvature singularities of our exact solution. The discussion in Sec.~\ref{sec:regularity} has shown that --- with the exception of the singular cases $c_3=\pm1$ --- the function $\ee^M$ cannot vanish in the Gowdy square $x\in[-1,1]$, $y\in[-1,1]$. However, it is still possible that there are zeros in the extended regions with $y\in\R$, which will be investigated in this section.

According to \eqref{eq:solnew1}, zeros of $\ee^M$ correspond to $U=V=0$, i.e. to
\begin{eqnarray}\label{eq:app1}
 U &=& c_3^2(1-x^2)(1-y)^3+4c_1^2(1+y)=0,\\
 V &=& 4c_1(1-y)[1-c_3x(2+y)]=0.\label{eq:app2}
\end{eqnarray}
For $c_3=0$ (the Taub case) we have $U=4c_1^2(1+y)$, $V=4c_1(1-y)$, so there are no simultaneous zeros of $U$ and $V$ (recall that $c_1>0$). This corresponds to the fact that the standard extensions of the Taub solution are free of curvature singularities. Therefore, we can now assume that $c_3>0$ (again taking advantage of the discrete symmetry of the solution). From $V=0$ we conclude that either $y=1$ or $y=\frac{1}{c_3 x}-2$ holds. In the former case, we obtain $U=8c_1^2\neq 0$, whereas the latter case leads to
\begin{equation}
 U=\frac{\omega(x)}{c_3 x^3},\quad
 \omega(x):=(1-x^2)(3c_3x-1)^3-4c_1^2x^2(c_3x-1).
\end{equation}
Hence the positions of curvature singularities are determined by those zeros of the fifths-degree polynomial $\omega(x)$ that are in the $x$-interval $[-1,1]$. For a discussion of these zeros we look at the following cases.

\begin{description}
 \item[1st case: $0<c_3<1$]\mbox{}\\
   From $\omega(0)=-1<0$, $\omega(1)=4c_1^2(1-c_3)>0$ and $\omega(\frac{1}{3c_3})=\frac{8c_1^2}{27c_3^2}>0$ we conclude that $\omega$ has a zero in the interval $x\in\Big(0,\min(1,\frac{1}{3c_3})\Big)$. Since, for $x<\frac{1}{3c_3}$, we have $y=\frac{1}{c_3x}-2>1$, this zero corresponds to a singularity in the \emph{past} extension. 
   Moreover, from $\omega(-1)=4c_1^2(1+c_3)>0$ and $\omega(0)=-1<0$ we conclude that $\omega$ has a zero in $(-1,0)$. For negative $x$ we get $y<-2$, i.e.\ we find a singularity in the \emph{future} extension.
   Due to $\omega(1)>0$ and $\lim_{x\to\infty}\omega(x)=-\infty$, there is another real zero in $[1,\infty)$. However, since our $x$-coordinate is restricted to $[-1,1]$, this zero has no physical meaning.
 \item[2nd case: $c_3=1$]\mbox{}\\
   In this singular case we see that $\omega(1)=0$, which corresponds to the singularity at $x=1$, $y=-1$ (point $C$ in Fig.~\ref{fig:Gowdy}). As in the first case, we also have a zero for $x\in\Big(0,\min(1,\frac{1}{3c_3})\Big)=(0,\frac13)$, corresponding to a singularity in the \emph{past} extension, and an irrelevant zero in $(1,\infty)$.
 \item[3rd case: $c_3>1$]\mbox{}\\
   As above, we observe that $\omega$ has a zero in $(0,\frac{1}{3c_3})$, corresponding to a singularity in the \emph{past} extension, and a zero in $(-1,0)$, corresponding to a singularity in the \emph{future} extension. In addition, we find a zero for $x\in(\frac{1}{c_3},1)$, because $\omega(\frac{1}{c_3})=8(1-\frac{1}{c_3^2})>0$ and $\omega(1)=-4c_1^2(c_3-1)<0$. This leads to a second singularity in the \emph{future} extension.
\end{description}
A detailed analysis of the polynomial $\omega$ and its corresponding Sturm's sequence reveals that $\omega$ has three real zeros and two complex zeros for all parameter values $c_1>0$, $c_3>0$. Therefore, there are no further real zeros besides the ones found in the above case-by-case analysis. 

Hence we arrive at the following result. The extended function $\ee^M$ always has one zero in the past extension and, depending on the value of $c_3$, either one or two zeros in the future extension. 

\section{Null geodesics on the axes\label{App2}}

In the following, we consider null geodesics that are restricted to either $\theta=0$ or $\theta=\pi$. Note that in the case of the spatially homogeneous Taub solution, one can find geodesics with $\theta=\mbox{constant}$ for arbitrary values of $\theta$. Here, however, due to the $\theta$-dependence of the metric potentials $M$ and $u$, only the special values $0$ and $\pi$ lead to geodesics.

We start by looking at the globally hyperbolic region $0<t<\pi$, before we study whether geodesics from that region can also be extended beyond the Cauchy horizons. Since we are interested in geodesics on the axes, and since the coordinates $(t,\theta,\rho_1,\rho_2)$ have a coordinate singularity there, we first introduce regular coordinates. In view of the intended extension of the geodesics through the Cauchy horizons, we start by introducing the coordinates $(x,y,\rho_1',\rho_2')$ with the coordinate transformation \eqref{eq:tran3}. In a next step we replace $\rho_1'$ and $\rho_2'$ with $\lambda_1'$ and $\lambda_2'$ via
\begin{equation}
 \lambda_1'=\frac{\rho_1'+\rho_2'}{2},\quad
 \lambda_2'=\frac{\rho_1'-\rho_2'}{2}.
\end{equation}
Finally, we remove the axes singularities. To this end, we separately study the cases $x=1$ and $x=-1$.

In a vicinity of the axis $x=1$, we locally introduce ``Cartesian coordinates'',
\begin{equation}
 p=\sqrt{1-x^2}\cos\lambda_2',\quad
 q=\sqrt{1-x^2}\sin\lambda_2',
\end{equation}
which replace $x$ and $\lambda_2$. The metric in terms of the new coordinates $(y,\lambda_1,p,q)$ is regular at the axis ($p=q=0$) and at the Cauchy horizons ($y=\pm1$). This can be verified with the explicit form of the solution and with the definitions of the constants $\kappa$, $\kappa_1$ and $\kappa_2$, which are introduced with the first of the above coordinate transformations, see Sec.~\ref{sec:extensions}. At the axis $x=1$, the metric has now the form
\begin{equation}
 x=1:\quad g = g_{yy}\,\dd y^2+2g_{y\lambda_1}\,\dd y\,\dd\lambda_1
    +g_{\lambda_1\lambda_1}\,\dd\lambda_1^2
    +\ee^M(\dd p^2+\dd q^2),
\end{equation}
where
\begin{eqnarray}
 \fl g_{yy} &=& -\frac{\ee^M}{1-y^2}+R_0(1-y^2)\ee^u\left(\frac{\kappa_1+\kappa_2}{1+y}-\frac{\kappa}{y-1}\right)^2,\\
 \fl g_{y\lambda_1'} &=& 2R_0(1-y^2)\ee^u\left(\frac{\kappa_1+\kappa_2}{1+y}-\frac{\kappa}{y-1}\right),\quad
 g_{\lambda_1'\lambda_1'}=4R_0(1-y^2)\ee^u.
\end{eqnarray}
We can find the geodesics by making use of the conservation laws that follow from the Killing vectors. The two Killing vectors $\partial_{\rho_1}$ and $\partial_{\rho_2}$ degenerate at $x=1$, where $\partial_{\rho_1}=\partial_{\rho_2}=\frac12\partial_{\lambda'_1}$. Hence we have only one conservation law, namely
\begin{equation}\label{eq:cons1}
 g(\partial_{\lambda_1'},v)=\mbox{constant}
\end{equation}
for the tangent vector $v^i=\dd x^i/\dd\lambda$ to the geodesic, where $\lambda$ is an affine parameter. Since $v^i$ is only determined up to a factor (we can rescale the null vector), we can set the constant to $2\eps R_0$, $\eps=\pm 1$.  Together with $g(v,v)=0$, we obtain the two equations
\begin{eqnarray}
 g_{\lambda_1'\lambda_1'}v^{\lambda_1}+g_{y\lambda_1'}v^y=2\eps R_0,\\
 g_{yy}(v^y)^2+2g_{y\lambda_1'}v^yv^{\lambda_1'}+g_{\lambda_1'\lambda_1'}(v^{\lambda_1'})^2=0,
\end{eqnarray}
which fix $v^y$ and $v^{\lambda_1'}$ (up to a sign). The remaining components $v^p$ and $v^q$ vanish since axis geodesics are characterized by $p=q=0$. In this way, we finally obtain
\begin{equation}\fl\label{eq:geo}
 v^y=-1,\quad
 v^{\lambda_1'}=\frac{\eps\ee^{-u}-2\kappa}{4(1-y)}+\frac{\eps\ee^{-u}+2(\kappa_1+\kappa_2)}{4(1+y)},\quad
 v^p=0,\quad
 v^q=0.
\end{equation}
Here, we have chosen a negative sign for $v^y$ in order to restrict to future-directed vectors. (Note that $y$ is \emph{decreasing} for increasing values of the time coordinate $t$.) 
We see that $-y$ can be used as an affine parameter. As a consequence, the geodesics are curves of the form 
\begin{equation}
 y=-\lambda,\quad
 \lambda_1'=\lambda_1'(\lambda),\quad
  p=0,\quad
  q=0, 
\end{equation}
where $\lambda_1'(\lambda)$ follows from $v^{\lambda_1'}$ in \eqref{eq:geo} with a $y$-integration. We denote the two classes of geodesics with $\eps=\pm 1$ as $\Gamma^{\pm}$.

For the calculation of geodesics on the second axis $\theta=\pi$, we can repeat the previous consideration, this time considering a vicinity of $x=-1$. The only difference is that ``Cartesian coordinates'' are now introduced via
\begin{equation}
 p=\sqrt{1-x^2}\cos\lambda_1',\quad
 q=\sqrt{1-x^2}\sin\lambda_1',
\end{equation}
i.e.\ this time we arrive at coordinates $(y,\lambda_2', p,q)$, where $\lambda_2'$ instead of $\lambda_1'$ is used as a regular coordinate. The geodesics turn out to be given by
\begin{equation}\fl
 v^y=-1,\quad
 v^{\lambda_2'}=\frac{\eps\ee^{-u}-2\kappa}{4(1-y)}+\frac{\eps\ee^{-u}+2(\kappa_1-\kappa_2)}{4(1+y)},\quad
 v^p=0,\quad
 v^q=0.
\end{equation}
Again we denote the geodesics with $\eps=\pm1$ as $\Gamma^{\pm}$.

For geodesics on either axis, we observe that the components of the tangent vector are analytic functions of $y$ in the interval $(-1,1)$, whereas they are potentially singular at $y=\pm 1$. More precisely, we have
\begin{eqnarray}
 x=1:\quad && v^{\lambda_1'}=\left\{
 \begin{array}{ll}
  \frac{(\eps-\mathrm{sgn}\,\kappa)c_1}{4(1-y)}+\mathcal O[(1-y)^0],& y\to1\\
  \frac{(\eps-\mathrm{sgn}\,\kappa_2)(1-c_3)^2}{4c_1(1+y)}+\mathcal O[(1+y)^0]  ,&y\to-1
 \end{array}\right.,\\
  x=-1:\quad && v^{\lambda_2'}=\left\{
 \begin{array}{ll}
  \frac{(\eps-\mathrm{sgn}\,\kappa)c_1}{4(1-y)}+\mathcal O[(1-y)^0],& y\to1\\
  \frac{(\eps-\mathrm{sgn}\,\kappa_2)(1+c_3)^2}{4c_1(1+y)}+\mathcal O[(1+y)^0]  ,&y\to-1
 \end{array}\right. .
\end{eqnarray}
We observe that there are \emph{no} singularities in $v^{i'}$, if we choose suitable signs for $\kappa$ and $\kappa_2$. Then, the components of $v^{i'}$ will be regular at and beyond $y=\pm 1$. This shows that there are extensions of our solutions in which the geodesics can be extended through the horizons. The ``correct'' sign choices can be summarized as follows:
\begin{enumerate}
 \item[(a)] An axis null geodesic $\Gamma^+$ extends through the past Cauchy horizon in $M^{ab}$ iff $a=+$, it extends through the future horizon in $M^{ab}$ iff $b=+$, and it extends through both horizons iff $a=b=+$.
 \item[(b)] An axis null geodesic $\Gamma^-$ extends through the past Cauchy horizon in $M^{ab}$ iff $a=-$, it extends through the future horizon in $M^{ab}$ iff $b=-$, and it extends through both horizons iff $a=b=-$.
\end{enumerate}
This should be compared with Lemma 3.2 in \cite{chrusciel93}, which states a similar result for the Taub solutions. However, whereas the Lemma in \cite{chrusciel93} considers more general null geodesics, our result applies only to axis geodesics.

\end{appendix}


\end{document}